\documentclass{article}

\usepackage{ra4}

\usepackage{etex}

\newcommand{\taba}{\phantom{\wedge}}

\usepackage{times}
\usepackage{mine}
\usepackage{yhmath}
\usepackage[colorinlistoftodos]{todonotes}
\usepackage{oz}
\usepackage{zed-csp}
\usepackage{graphicx}

\usepackage[T1]{fontenc}

\usepackage{mathrsfs}
\usepackage{rcp}
\usepackage{color}		
\usepackage{epsfig}
\usepackage{stmaryrd}

\def \hasgn{\asgn}
\def\cdota{\!\cdot\!}
\def \Eval{{\sf eval}} 
 
\def \Update{{\sf update}}

\def \reachable{{\sf RE}}

\newcommand{\AG}[1]{\aang{#1}} 
\newcommand{\Idle}{{\sf Idle}}
\newcommand{\FinIdle}{{\sf fin\_Idle}}
\newcommand{\InfIdle}{{\sf inf\_Idle}}
 
\newcommand{\Fin}{{\sf fin}}
\newcommand{\Inf}{{\sf inf}}

\renewcommand{\qed}{\ensuremath{{}_\Box}}
\newcommand{\lub}{{\sf lub}}
\newcommand{\glb}{{\sf glb}}

\newcommand{\st}{~{\scriptscriptstyle ^\bullet}~}

\def\figrule{\rule{\columnwidth}{0.5pt}}

\def \rely {\mathop{\textsc{Rely}}}
\def \Init {\mathop{\textsc{Init}}}

\def \enf {\mathop{\textsc{Enf}}}

\newcommand{\NoteEnv}[3]{\newenvironment{#1}{\color{#3}#2 }{}}
\definecolor{brijeshcolor}{rgb}{1,0.2,0}
\definecolor{johncolor}{cmyk}{1,0.3,0.4,0.3}

\NoteEnv{brijesh}{Brijesh says:}{brijeshcolor}
\NoteEnv{modified}{}{blue}
\NoteEnv{ian}{Ian}{iancolor}
\NoteEnv{lindsay}{Lindsay}{lindsaycolor}
\NoteEnv{john}{John}{johncolor}

\DeclareMathSymbol{\Diamond}{\mathord}{lasy}{"33}

\usepackage{amsthm}
\theoremstyle{plain}
\newcounter{thm}
\newtheorem{theorem}{Theorem}[section]
\newtheorem{lemma}[thm]{Lemma}

\theoremstyle{definition}
\newtheorem{definition}{Definition}[section]

\newtheorem{example}{Example}[section]

\def \llb {\llbracket}
\def \rrb {\rrbracket}
\def\abssynt{\mathop{:\joinrel:\joinrel=}}

\newcommand{\Context}[2]
{\!\llb #1 
    \begin{array}[c]{@{~~}|@{~~}l@{}}
      #2
    \end{array}\rrb }

\newcommand{\prev}{\varominus}

\newcommand{\Rely}[2]
{\rely #1 \st
    \begin{array}[c]{@{}l@{}}
      #2
    \end{array}}

\newcommand{\Enf}[2]
{\enf #1 \st
    \begin{array}[c]{@{}l@{}}
      #2
    \end{array}}

\makeatletter

\newcommand*\wthelper[2]{%
        \hbox{\dimen@\accentfontxheight#1%
                \accentfontxheight#11.5\dimen@
                $\m@th#1\widetilde{#2}$%
                \accentfontxheight#1\dimen@
        }%
}

\newcommand*\accentfontxheight[1]{%
        \fontdimen5\ifx#1\displaystyle
                \textfont
        \else\ifx#1\textstyle
                \textfont
        \else\ifx#1\scriptstyle
                \scriptfont
        \else
                \scriptscriptfont
        \fi\fi\fi3
}
\makeatother

\newcommand{\Par}{\textstyle\mathop{\|}}

\newcommand{\Empty}{{\sf empty}}
\newcommand{\Always}{\textstyle\mathord{\boxdot}}
\newcommand{\Sometime}{\raisebox{0.1em}{\rotatebox[origin=c]{45}{$\textstyle\boxdot$}}}
\newcommand{\Def}{\textstyle\mathord{\boxast}}
\newcommand{\Pos}{\raisebox{0.1em}{\rotatebox[origin=c]{45}{$\textstyle\boxast$}}}

\newcommand{\Defp}{\mathop{\Def_p}}
\newcommand{\Posp}{\mathop{\Pos\!_p}}

\def\adjoins{\mathbin{\varpropto}}

\def\ch{\mathbin{;}}

\def \kif{\mathop{\mathsf{if}}}
\def \kthen{\mathbin{\mathsf{then}}}
\def \kelse{\mathbin{\mathsf{else}}}

\def \klet {\mathop{{\sf let}}}
\def \kin {\mathbin{{\sf in}}}

\def \bs {\backslash}

\def \deref{\mathop{*}}

\def \seq {\mathrm{seq}}

\title{Simplifying proofs of linearisability using layers of
  abstraction} 

\author{Brijesh Dongol\ and\ John Derrick \\
  \\
  \small{Department of Computer Science} \\
  \small{The University of Sheffield, S1 4DP, UK} \\
  \small{\texttt{B.Dongol@sheffield.ac.uk, J.Derrick@dcs.shef.ac.uk}}}

\date{}

\begin{document}

\maketitle

\begin{abstract}
  Linearisability has become the standard correctness criterion for
  concurrent data structures, ensuring that every history of
  invocations and responses of concurrent operations has a matching
  sequential history. Existing proofs of linearisability require one
  to identify so-called linearisation points within the operations
  under consideration, which are atomic statements whose execution
  caus\-es the effect of an operation to be felt. However,
  identification of linearisation points is a non-trivial task,
  requiring a high degree of expertise. For sophisticated algorithms
  such as Heller et al's lazy set, it even is possible for an
  operation to be linearised by the concurrent execution of a
  statement outside the operation being verified. This paper
  proposes an alternative method for verifying linearisability that
  does not require identification of linearisation points. Instead,
  using an interval-based logic, we show that every behaviour of
  each concrete operation over any interval is a possible behaviour
  of a corresponding abstraction that executes with coarse-grained
  atomicity. This approach is applied to Heller et al's lazy set to
  show that verification of linearisability is possible without
  having to consider linearisation points within the program code.
\end{abstract}

\section{Introduction}
\label{sec:intro}
Development of correct fine-grained concurrent data structures has
received an increasing amount of attention over the past few years as
the popularity of multi/many-core architectures has increased. An
important correctness criterion for such data structures is
\emph{linearisability} \cite{Herlihy90}, which guarantees that every
history of invocations and responses of the concurrent operations on
the data structure can be rearranged without violating the ordering
within a process such that the rearranged history is a valid
sequential history. % Linearisability guarantees
% the existence of a so-called \emph{linearisation point} for each
% operation, which is an atomic statement whose execution causes the
% effect of the operation to take place, i.e., the execution of the
% linearisation point in the concrete program causes a corresponding
% effect in its abstract representation.  As a result, the proof
% techniques that have been developed over the years are based on the
% identification of linearisation points. 
A number of proof techniques developed over the years match concurrent
and sequential histories by identifying an atomic \emph{linearising
  statement} within the concrete code of each operation, whose
execution corresponds to the effect of the operation taking place.
However, due to the subtlety and complexity of concurrent data
structures, identification of linearising statements within the
concrete code is a non-trivial task, and it is even possible for an
operation to be linearised by the execution of other concurrent
operations. An example of such behaviour occurs in Heller et al's lazy
set algorithm, which implements a set as a sorted linked list
\cite{HHLMSS07} (see \reffig{fig:lazyset}). In particular, its {\tt
  contains} operation may be linearised by the execution of a
concurrent {\tt add} or {\tt remove} operation and the precise
location of the linearisation point is dependent on how much of the
list has been traversed by the {\tt contains} operation.  This paper
presents a method for simplifying proofs of linearisability using
Heller et al's lazy set as an example.

% As a result, verifying linearisability of {\tt contains} using
% linearisation points requires the use of sophisticated proof
% techniques such as backwards simulation \cite{CGLM06} and non-atomic
% refinement \cite{DSW11}.

An early attempt at verifying linearisability of Heller et al's lazy
set is that of Vafeiadis et al, who extend each linearising statement
with code corresponding to the execution of the abstract operation so
that execution of a linearising statement causes the corresponding
abstract operation to be executed \cite{VHHS06}. However, this
technique is incomplete and cannot be used to verify the {\tt
  contains} operation, and hence, its correctness is only treated
informally \cite{VHHS06}. These difficulties reappear in more recent
techniques: ``In [Heller et al's lazy set] algorithm, the correct
abstraction map lies outside of the abstract domain of our
implementation and, hence, was not found.''  \cite{Vaf10}.  The first
complete linearisability proof of the lazy set was given by Colvin et
al \cite{CGLM06}, who map the concrete program to an abstract set
representation using simulation to prove data refinement. To verify
the {\tt contains} operation, a combination of forwards and backwards
simulation is used, % \footnote{Forwards simulation alone is known to be
  % incomplete \cite{deRoever98}. This fact is closely related to
  % incompleteness of the methods proposed in \cite{VHHS06,Vaf10} ---
  % algorithms that require backwards simulation are precisely those
  % that cause problems with the methods in \cite{VHHS06,Vaf10}.}, 
which
involves the development of an intermediate program $IP$ such that
there is a backwards simulation from the abstract representation to
$IP$, and a forwards simulation from $IP$ to the concrete
program. More recently, O'Hearn et al use a so-called hindsight lemma
(related to backwards simulation) to verify a variant of Heller's lazy
set algorithm \cite{OHea10}.  Derrick et al use a method based on
\emph{non-atomic} refinement, which allows a single atomic step of the
concrete program to be mapped to several steps of the abstract
\cite{DSW11}.

Application of the proof methods in \cite{VHHS06,CGLM06,OHea10,DSW11}
remains difficult because one must acquire a high degree of expertise
of the program being verified to correctly identify its linearising
statements. For complicated proofs, it is difficult to determine
whether the implementation is erroneous or the linearising statements
have been incorrectly chosen. Hence, we propose an approach that
eliminates the need for identification of linearising statements in
the concrete code by establishing a refinement between the
fine-grained implementation and an abstraction that executes with
coarse-grained atomicity \cite{DD12}. The idea of mapping fine-grained
programs to a coarse-grained abstraction has been proposed by Groves
\cite{Gro08} and separately Elmas et al \cite{EQSST10}, where the
refinements are justified using \emph{reduction}
\cite{Lip75}. However, unlike our approach, their methods must
consider each pair of interleavings, and hence, are not
compositional. Turon and Wand present a method of abstraction in a
compositional rely/guarantee framework with separation logic
\cite{TW11}, but only verify a stack algorithm that does not require
backwards reasoning.

% establishes a refinement between a fine-grained implementation and an
% abstraction that executes with coarse-grained atomicity. The
% refinement proof does not require identification of linearising
% statements in the implementation . Our method only requires
% verification of linearisability of an abstraction (see
% \reffig{fig:steps}). Due to the coarse-granularity of its atomic
% statements, the linearising statements of the abstraction are
% straightforward to identify.

Capturing the behaviour of a program over its interval of execution is
crucial to proving linearisability of concurrent data structures. In
fact, as Colvin et al point out: ``The key to proving that [Heller et
al's] lazy set is linearisable is to show that, for any failed {\tt
  contains(x)} operation, {\tt x} is absent from the set at some point
during its execution.''  \cite{CGLM06}.  Hence, it seems
counter-intuitive to use logics that are only able to refer to the pre
and post states of each statement (as done in
\cite{VHHS06,CGLM06,DSW11,Vaf10}). Instead, we use a framework based
on \cite{DDH12} that allows reasoning about the fine-grained atomicity
of pointer-based programs over their intervals of execution. By
considering complete intervals, i.e., those that cover both the
invocation and response of an operation, one is able to determine the
future behaviour of a program, and hence, backwards reasoning can
often be avoided. % This is in contrast to the proofs in
% \cite{VHHS06,CGLM06,DSW11} that match each step of execution of a
% concrete program to an abstract execution. 
For example, B{\"a}umler et al \cite{BSTR11} use an interval-based
approach to verify a lock-free queue without resorting to backwards
reasoning, as is required by frameworks that only consider the
pre/post states of a statement \cite{DGLM04}. However, unlike our
approach, B{\"a}umler et al must identify the linearising statements
in the concrete program, which is a non-trivial step.

An important difference between our framework and those mentioned
above is that we assume a truly concurrent execution model and only
require interleaving for conflicting memory accesses
\cite{DD12,DDH12}. Each of the other frameworks mentioned above assume
a strict interleaving between program statements. Thus, our approach
captures the behaviour of program in a multicore/multiprocesor
architecture more faithfully.  

The main contribution of this paper is the use of the techniques in
\cite{DD12} to simplify verification of a complex set algorithm by
Heller et al. This algorithm presents a challenge for linearisability
because the linearisation point of the {\tt contains} operation is
potentially outside the operation itself \cite{DSW11}. We propose a
method in which the proof is split into several layers of abstraction
so that linearisation points of the fine-grained implementation need
not be identified. As summarised in \reffig{fig:steps}, one must
additionally prove that the coarse-grained abstraction is
linearisable, however, due to the coarse granularity of atomicity, the
linearising statements are straightforward to identify and the
linearisability proof itself is simpler \cite{DD12}. Other
contributions of this paper include a method for reasoning about truly
concurrent program executions and an extension of the framework in
\cite{DDH12} to enable reasoning about pointer-based programs, which
includes methods for reasoning about expressions non-deterministically
\cite{HBDJ13}.

% We present Heller et al's lazy set algorithm in
% \refsec{sec:list-based-conc}.  Our methodology integrates a number
% approaches, namely an interval-based framework (see
% \refsec{sec:an-interval-based}) with methods for reasoning about the
% actual and apparent states within an interval, as well as permissions.
% \refsec{sec:behaviour-refinement-1} then presents the refinement
% theory.  This integrated theory is applied to verify part of the lazy
% set algorithm in \refsec{sec:verif-lazy-set}.

% \begin{brijesh}
%   A novel interval-based framework for reasoning about pointer-based
%   programs. Pointers may be nested and ...
% \end{brijesh}

\section{A list-based concurrent set}  
\label{sec:list-based-conc}
\begin{figure}[t]
  \centering\small

  \fbox{\begin{minipage}[t]{.3\textwidth}
    \tt 
  add(x): 
    
  \ A1: n1, n3:= 

    \vspace{-0.3em}    
    \ \ \ \ \ \ \ \ locate(x); 

    \ A2: {\bf if} n3.val != x 
    % \ \ \ \ {\bf then} 

    \ A3: \ \ n2:= 

    \vspace{-0.3em}    
    \ \ \ \ \ \ \ \ \ {\bf new} Node(x); 

    \ A4: \ \ n2.nxt := n3; 

    \ A5: \ \ n1.nxt := n2; 

    \ A6: \ \ res := true 

    \ A7: {\bf else} res := false 
    
    \ \ \ \ \ {\bf endif};

    \ A8: n1.unlock(); 

    \ A9: n3.unlock(); 

    A10:\ {\bf return} res
  \end{minipage}}
\fbox{
  \begin{minipage}[t]{.3\textwidth}
    \tt 
    remove(x): 
    
    \ R1:\  n1, n2 := 

    \vspace{-0.3em}    
    \ \ \ \ \ \ \ \ locate(x); 

    \ R2:\  {\bf if} n2.val = x %{\bf then}

    \ R3:\ \ \ n2.mrk := true;

    \ R4:\  \ \ n3 := n2.nxt;

    \ R5:\  \ \ n1.nxt := n3;

    \ R6:\  \ \ res := true

    \ R7:\ {\bf else} res := false 

    \ \ \ \ \ {\bf endif};

    \ R8:\ n1.unlock();

    \ R9:\ n2.unlock();

    R10:\ {\bf return} res    
    \

    \vspace{-.3em}
    \
  \end{minipage}
  }
  \fbox{
  \begin{minipage}[t]{.3\textwidth}
    \tt 
    contains(x): 
    
    C1: n1 := Head; 

    C2: {\bf while} (n1.val < x) %{\bf do}

    C3: \ \ n1 := n1.nxt
    
    \ \ \ \ {\bf enddo};

    C4: res :=  (n1.val = x) 

    \vspace{-0.3em}
    \ \ \ \ \ \ \ \ \ \ \ {\bf and} !n1.mrk 
    
    C5: {\bf return} res
    \ 
    
    \

    \ 
    
    \
    
    \
     
    \

    \vspace{-.3em}
    \
  \end{minipage}
  }
  
  \fbox{
    \begin{minipage}[t]{0.475\textwidth}
    \tt 
    locate(x):

    \ \ \ \ \ {\bf while} (true) {\bf do}
    
    \ L1:\ \ \ pred := Head; 

    \ L2:\ \ \ curr := pred.nxt; 

    \ L3:\ \ \ {\bf while} (curr.val < x) {\bf do} 

    \ L4:\ \ \ \ \ pred := curr; 

    \ L5:\ \ \ \ \ curr := pred.nxt {\bf enddo};

    \ L6:\ \ \ pred.lock();
  \end{minipage}
    \begin{minipage}[t]{0.47\textwidth}
    \tt 
    \
    
    \

    \ L7:\ \ \ curr.lock();

    \ L8:\ \ \ {\bf if} !pred.mrk 
    {\bf and} !curr.mrk 

    \ \ \ \ \ \ \ \ \ \ 
    {\bf and} pred.nxt = curr
    
    \ L9:\ \ \ \ \ {\bf return} pred, curr

    L10:\ \ \ {\bf else} pred.unlock(); 
    
    L11:\ \ \ \ \ curr.unlock() {\bf endif} {\bf enddo}
  \end{minipage}
  }
  \caption{Heller et al's lazy set algorithm}
  \label{fig:lazyset}
\end{figure}

Heller et al \cite{HHLMSS07} implement a set as a concurrent algorithm
operating on a shared data structure (see \reffig{fig:lazyset}) with
operations {\tt add} and {\tt remove} to insert and delete elements
from the set, and an operation {\tt contains} to check whether an
element is in the set. The concurrent implementation uses a shared
linked list of node objects with fields $val, nxt, mrk$, and $lck$,
where $val$ stores the value of the node, $nxt$ is a pointer to the
next node in the list, $mrk$ denotes the marked bit and $lck$ stores
the identifier of the process that currently holds the lock to the
node (if any) \cite{HHLMSS07}.  The list is sorted in strictly
ascending values order (including marked nodes).

Operation {\tt locate(x)} is used to obtain pointers to two nodes
whose values may be used to determine whether or not {\tt x} is in the
list --- the value of the predecessor node {\tt pred} must always be
less than {\tt x}, and the value of the current node {\tt curr} may
either be greater than {\tt x} (if {\tt x} is not in the list) or
equal to {\tt x} (if {\tt x} is in the list). Operation {\tt add(x)}
calls {\tt locate(x)}, then if {\tt x} is not already in the list
(i.e., value of the current node {\tt n3} is strictly greater than
{\tt x}), a new node {\tt n2} with value field {\tt x} is inserted
into the list between {\tt n1} and {\tt n3} and {\tt true} is
returned. If {\tt x} is already in the list, the {\tt add(x)}
operation does nothing and returns {\tt false}. Operation {\tt
  remove(x)} also starts by calling {\tt locate(x)}, then if {\tt x}
is in the list the current node {\tt n2} is removed and {\tt true} is
returned to indicate that {\tt x} was found and removed. If {\tt x} is
not in the list, the {\tt remove} operation does nothing and returns
{\tt false}. Note that operation {\tt remove(x)} distinguishes between
a logical removal, which sets the marked field of {\tt n2} (the node
corresponding to {\tt x}), and a physical removal, which updates the
{\tt nxt} field of {\tt n1} so that {\tt n2} is no longer
reachable. Operation {\tt contains(x)} iterates through the list and
if a node with value greater or equal to {\tt x} is found, it returns
{\tt true} if the node is unmarked and its value is equal to {\tt x},
otherwise returns {\tt false}.
% The program distinguishes between a logical
% removal of {\tt x} (which sets the $mrk$ field of the node
% corresponding to {\tt x} to {\tt false}), and a physical removal
% (where the $nxt$ fields of the nodes are updated so the node
% corresponding to {\tt x} can no longer be reached). Consequently
% A consequence of this distinction is that the {\tt contains(x)}
% operation becomes more complicated, because it must check that the
% node corresponding to {\tt x} is both logically and physically in the
% list. 
% In this paper, we focus on a proof of the {\tt contains(x)}
% operation \cite{VHHS06,CGLM06,DSW11}.
% \begin{brijesh}
%   Can use Vafeiadis to prove lin of coarse-grained abstraction.
% \end{brijesh}

% , e.g., ``The issue is that it is not
% possible to statically determine the linearisation point of {\tt
%   contains} as it depends on future behaviour of processes other than
% the one currently executing {\tt contains}.''
% This paper aims to
\begin{figure}[!t]
  \begin{minipage}[b]{0.65\textwidth}
    \scalebox{0.8}{\input{exec.pspdftex}}      
    \caption{Execution of {\tt contains(x)} over $\Delta_p$ that
      returns $true$}
    \label{fig:tf}
  \end{minipage}
  \hfill
  \begin{minipage}[b]{0.32\textwidth}
    \scalebox{0.85}{\input{steps.pspdftex}}
    \caption{Proof steps}
    \label{fig:steps}
  \end{minipage}
\end{figure}

The complete specification consists of a number of processes, each of
which may execute its operation on the shared data structure. For the
concrete implementation, therefore, the set operations can be executed
concurrently by a number of processes, and hence, the intervals in
which the different operations execute may overlap. Our basic
semantic model uses \emph{interval predicates} (see
\refsec{sec:an-interval-based}), which allows formalisation of a
program's behaviour with respect to an \emph{interval} (which is a
contiguous set of times), and an infinite \emph{stream} (that maps each
time to a state). For example, consider \reffig{fig:tf}, which depicts
an execution of the lazy set over interval $\Delta$ in stream $s$, a
process $p$ that executes a {\tt contains(x)} that returns $true$ over
$\Delta_p$, a process $q$ that executes {\tt remove(x)} and {\tt
  add(y)} over intervals $\Delta_q$ and $\Delta_q'$, respectively, and
a process $u$ that executes {\tt add(x)} over interval
$\Delta_u$. Hence, the shared data structure may be changing over
$\Delta_p$ while process $p$ is checking to see whether $x$ is in the
set.

Correctness of such concurrent executions is judged with respect to
\emph{linearisability}, the crux of which requires the existence of an
atomic \emph{linearisation point} within each interval of an
operation's execution, corresponding to the point at which the effect
of the operation takes place \cite{Herlihy90}. The ordering of
linearisation points defines a sequential ordering of the concurrent
operations and linearisability requires that this sequential ordering
is valid with respect to the data structure being implemented.  For
the execution in \reffig{fig:tf}, assuming that the set is initially
empty, because {\tt contains(x)} returns $true$, a valid linearisation
corresponds to a sequential execution $Seq_1 \sdef \texttt{add(x);
  contains(x); remove(x); add(y)}$ obtained by picking linearisation
points within $\Delta_u$, $\Delta_p$, $\Delta_q$ and $\Delta_q'$ in
order. Note that a single concurrent history may be linearised by more
than one valid sequential history, e.g., the execution in
\reffig{fig:tf} can correspond to the sequential execution $Seq_2
\sdef \texttt{remove(x); add(x); contains(x); add(y)}$. The abstract
sets after completion of $Seq_1$ and $Seq_2$ are $\{y\}$ and
$\{x,y\}$, respectively. Unlike $Seq_1$, operation {\tt remove(x)} in
$Seq_2$ returns $false$. Note that a linearisation of $\Delta_q'$
cannot occur before $\Delta_q$ because {\tt remove(x)} responds before
the invocation of {\tt add(y)}.

Herlihy and Wing formalise linearisability in terms of histories of
invocation and response events of the operations on the data structure
in question \cite{Herlihy90}. Clearly, reasoning about such histories
directly is infeasible, and hence, existing methods (e.g.,
\cite{CGLM06,DSW11,VHHS06}) prove linearisability by identifying an
atomic {\em linearising statement} within the operation being verified
and showing that this statement can be mapped to the execution of a
corresponding abstract operation.  However, due to the fine
granularity of the atomicity and inherent non-determinism of
concurrent algorithms, identification of such a statement is
difficult. The linearising statement for some operations may actually
be outside the operation, e.g., none of the statements {\tt C1}-{\tt
  C5} are valid linearising statements of {\tt contains(x)}; instead
{\tt contains(x)} is linearised by the execution of a statement within
{\tt add(x)} or {\tt remove(x)} \cite{DSW11}.
% may start an enq operation (say doing E0 and E1)
  % but then another process may execute its own atomic step (e.g.,
  % start a deq). 
  % Verification that the concrete implementation is
  % somehow correct with respect to abstract, atomic enqueue and dequeue
  % operations is the crux of the problem and linearisability is the
  % proof obligation.

As summarised in \reffig{fig:steps}, we decompose proofs of
linearisability into two steps, the first of which proves that a
fine-grained implementation refines a program that executes the same
operations but with coarse-grained atomicity. The second step of the
proof is to show that the abstraction is linearisable. The atomicity
of a coarse-grained abstraction cannot be guaranteed in hardware
(without the use of contention inducing locks), however, its
linearisability proof is much simpler \cite{DDH12}. Because we prove
behaviour refinement, any behaviour of the fine-grained implementation
is a possible behaviour of the coarse-grained abstraction, and hence,
an implementation is linearisable whenever the abstraction is
linearisable. Our technique does not require identification of the
linearising statements in the implementation.

A possible coarse-grained abstraction of {\tt contains(x)} is an
operation that is able to test whether {\tt x} is in the set in a
single atomic step (see \reffig{fig:labs}), unlike the implementation
in \reffig{fig:lazyset}, which uses a sequence of atomic steps to
iterate through the list to search for a node with value {\tt x}.
Therefore, as depicted in \reffig{fig:tf}, an execution of {\tt
  contains} that returns $true$, i.e., $\texttt{C1} \ch (\texttt{C2}
\ch \texttt{C3})^\omega \ch \texttt{C4} \ch {\bf return}\ true$, is
required to refine a coarse-grained abstraction $\aang{x \in absSet}
\ch {\bf return}\ true$, where {\tt C1} - {\tt C4} are the labels of
{\tt contains} in \reffig{fig:lazyset} and $\aang{x \in absSet}$ is a
guard that is atomically able to test whether $x$ is in the abstract
set.  In particular, $\aang{x \in absSet}$ holds in an interval
$\Omega$ and stream $s$ iff there is a time $t$ in $\Omega$ such that
$x \in absSet.(s.t)$. Streams are formalised in
\refsec{sec:interval-predicates}. Note that both $\aang{x \in absSet}$
and $\aang{x \notin absSet}$ may hold within $\Delta_p$; the
refinement in \reffig{fig:tf} would only be invalid if for all $t \in
\Delta_p$, $x \notin absSet.(s.t)$ holds.

Proving refinement between a coarse-grained abstraction and an
implementation is non-trivial due to the execution of other
(interfering) concurrent processes. Furthermore, our execution model
allows non-conflicting statements (e.g., concurrent writes to
different locations) to be executed in a truly concurrent manner. We
use compositional rely/guarantee-style reasoning \cite{Jon83} to
formalise the behaviour of the environment of a process and allow the
execution of an arbitrary number of processes in the environment. Note
that unlike Jones \cite{Jon83}, who assumes rely conditions are
two-state relations, rely conditions in our framework are interval
predicates that are able to refer to an arbitrary number of states.
 
% \begin{brijesh}
%   \begin{itemize}
%   % \item In this paper, we focus on a proof of the contains operation.
%   % \item Linearisability ensures that every concurrent operations that
%   % \item As we have mentioned, for any complete execution of contains,
%   %   there must be a state $\sigma$ such that the abstract set
%   %   corresponding to $\sigma$
%   \item How is ITL used to do the reasoning that we need
%     \begin{itemize}
%     \item ITL (basic unifying semantic framework)
%     \item Permissions (concurrency)
%     \item Apparent/Actual (fine-grained atomicity)
%     \end{itemize}
%   \end{itemize}
% \end{brijesh}

% To prove the refinement in \reffig{fig:tf}, we use interval predicates
% to model the behaviour of a program over time, providing our basic
% semantic framework \cite{DDH12,DH12MPC,DHMS12}. Hence, our framework
% includes fractional permissions to control access to shared state
% \cite{Boy03,DD12}. To further exploit our interval-based framework,
% for reasoning about the states that are apparent to a process, which
% accommodates for the fact that expression evaluation takes multiple
% steps, and that there may be interference \cite{HBDJ13}.

\section{Interval-based framework}
\label{sec:an-interval-based}

To simplify reasoning about the linked list structure of the lazy
list, the domain of each state distinguishes between variables and
addresses. We use a language with an abstract syntax that closely
resembles program code, and use interval predicates to formalise
interval-based behaviour. Fractional permissions are used to control
conflicting accesses to shared locations.

\paragraph{Commands.}
We assume variable names are taken from the set $Var$, values have
type $Val$, addresses have type $Addr \sdef \nat$, $Var \cap Addr =
\emptyset$ and $Addr \subseteq Val$. A \emph{state} over $VA \subseteq
Var \cup Addr$ has type $State_{VA} \sdef VA \fun Val$
and a \emph{state predicate} has type $State_{VA} \fun \bool$.

The objects of a data structure may contain fields, which we assume
are of type $Field$. We assume that every object with $m$ fields is
assigned $m$ contiguous blocks of memory and use $offset : Field \fun
\nat$ to obtain the offset of $f \in Field$ within this block
\cite{Vaf07}, e.g., for the fields of a node object, we assume that
$offset.val = 0$, $offset.nxt = 1$, $offset.mrk = 2$ and $offset.lck =
3$.

We assume the existence of a function $eval$ that evaluates a given
expression in a given state. The full details of expression evaluation
are elided. To simplify modelling of pointer-based programs, for an
address-valued expression $ae$, we introduce expressions $\deref ae$,
which returns the value at address $ae$, $ae \cdota f$, which returns
the address of $f$ with respect to $ae$. For a state $\sigma$, we
define $eval.(\deref ae).\sigma \sdef \sigma.(eval.ae.\sigma)$ and
$(ae \cdota f).\sigma \sdef eval.ae.\sigma + offset.f$. We also define
shorthand $ae \mapsto f \sdef \deref (ae \cdota f) $, which returns
the value at $ae \cdota f$ in state $\sigma$.

\begin{figure}[!t]
  \centering
  \figrule \small

  $\begin{array}{@{}rcl@{}}
    \mathsf{CLoop}(p,x) & \sdef &  ([(n1_p\mapsto val)< x]
    \ch n1_p \asgn (n1_p\mapsto nxt))^\omega \ch 
    {[}(n1_p\mapsto val) \geq x{]}
     \\
    \mathsf{Contains}(p,x) & \sdef &
    \begin{array}[t]{@{}l@{}}
      cl_1: n1_p \asgn Head \ch cl_2:  \mathsf{CLoop}(p,x) \ch {} \\
      cl_3: res_p \asgn (\neg
      (n1_p\mapsto mrk) \land 
      (n1_p\mapsto val) = x)\\
      % \left(\begin{array}[c]{@{}l@{}}
      %   C4t:
      %   [n1_p\mapsto mrk] \ch C5: res_p \asgn false \\
      %   \sqcap \\
      %   C4f: [\neg (n1_p\mapsto mrk)]
      %   \ch C6: res_p \asgn ((curr\mapsto val) = x)
      % \end{array}\right)
    \end{array}
  \end{array}$
  \smallskip
  
  $\begin{array}[t]{@{}rcl@{}} 
    HTInit & \sdef & (Head \longmapsto (-\infty, Tail, false, null))
    \land (Tail \longmapsto
    (\infty, null, false, null))
    \smallskip \\
    \mathsf{S}(p) & \sdef & 
    \Context{n1_p,n2_p,n3_p,res_p}{(\bigsqcap_{x : \integer}\
      \mathsf{Add}(p,x) \sqcap \mathsf{Remove}(p,x) \sqcap
      \mathsf{Contains}(p,x))^\omega}
    \smallskip\\
    \mathsf{Set}(P) & \sdef & \Context{Head, Tail}{\rely \ola{HTInit}
      \st \Par_{p:P}\ \mathsf{S}(p)}
  \end{array}$
  \figrule
  \caption{Formal model of the lazy set operations}
  \label{fig:formal-lazyset}
\end{figure}

Assuming that $Proc$ denotes the set of process ids, for a set of
variables $Z$, state predicate $c$, variable or address-valued
expression $vae$, expression $e$, label $l$, and set of processes $P
\subseteq Proc$, the abstract syntax of a command is given by $Cmd$
below, where $C, C_1, C_2, C_p \in Cmd$.
$$
\begin{array}[t]{@{}rcl@{}}
  Cmd & \abssynt & 
  \begin{array}[t]{@{}l@{}}
    % \Chaos ~~\mid ~~ 
    \Idle ~~\mid ~~ [c] ~~\mid~~ \AG{c} ~~\mid~~
    vae \hasgn e ~~ \mid~~
    C_1 \ch C_2 ~~\mid~~ 
    C_1 \sqcap C_2
    ~~\mid~~  
    C^\omega 
    ~~\mid~~ \Par_{p:P}\ C_p 
    ~~\mid~~ \Context{Z}{C} ~~\mid~~ l : C 
  \end{array}
\end{array}$$
Hence a command is either % a $\Chaos$ command, 
$\Idle$, a
guard $[c]$, an atomically evaluated guard $\AG{c}$, an assignment $vae \asgn e$, a
sequential composition $C_1 \ch 
C_2$, a non-deterministic choice $C_1 \sqcap C_2$, a possibly infinite
iteration $C^\omega$, a parallel composition $\Par_{p:P}\ C_p $, a
command $C$ within a context $Z$ (denoted $\Context{Z}{C}$), or a labelled 
command $l : C$. % The interval-based semantics of commands are
% formalised in \reffig{fig:beh-def}. 
 
A formalisation of part of Heller et al's lazy list using the syntax
above is given in \reffig{fig:formal-lazyset}, where $P \subseteq
Proc$. Operations {\tt add(x)}, {\tt remove(x)} and {\tt contains(x)}
executed by process $p$ are modelled by commands $\mathsf{Add}(p,x)$,
$\mathsf{Remove}(p,x)$ and $\mathsf{Contains}(p,x)$, respectively. We
assume that $n \longmapsto (vv, nn, mm, ll)$ denotes $(n \mapsto val =
vv) \land (n \mapsto nxt = nn) \land (n \mapsto mrk = mm) \land (n
\mapsto lck = ll)$. Details of $\mathsf{Add}(p,x)$ and
$\mathsf{Remove}(p,x)$ are elided and the $\rely$ construct is
formalised in \refsec{sec:behaviour-refinement-1}.\footnote{The
  formalisation is given in Appendix A.}  Note that unlike the methods
in \cite{CGLM06,DSW11}, where labels identify the atomicity, we use
labels to simplify formalisation of the rely conditions of each
process, and may correspond to a number of atomic steps.
Furthermore, % (see
% \refsec{sec:semantics})
guard evaluation is formalised with respect to the set of states
apparent to a process (see \refsec{sec:eval-state-pred}), and hence,
unlike \cite{VHHS06,CGLM06,DSW11}, we need not split complex
expressions into their atomic components. % Hence our model of the
% lazy set is more general than, where a specific ordering of execution
% must be chosen. 
For example, in \cite{VHHS06,CGLM06,DSW11}, the expression at {\tt C4}
(\reffig{fig:lazyset}) must be split into two expressions {\tt
  curr.val = x} and {\tt !curr.mrk} to explicitly model the fact that
interference may occur between accesses to {\tt curr.val} and {\tt
  curr.mrk}. % Further details of our execution model and its

\paragraph{Interval predicates.}
\label{sec:interval-predicates}
A (discrete) {\em interval} (of type $Intv$) is a contiguous set of
time (of type $Time \sdef \integer$), i.e., $Intv \sdef \{\Delta
\subseteq Time \mid \all t, t' : \Delta \st \all u : Time @ t \leq u
\leq t' \imp u \in \Delta\}$.  Using `.' for function application, we
let $\lub.\Delta$ and $\glb.\Delta$ denote the \emph{least upper} and
\emph{greatest lower} bounds of an interval $\Delta$, respectively,
where $\lub.\emptyset \sdef -\infty$ and $\glb.\emptyset \sdef
\infty$. % If the size of $\Delta$ is infinite and $\glb.\Delta \in
% \integer$, then $\lub.\Delta = \infty$ (i.e., is not a member of
% $\integer$) and if $\Delta$ is infinite and $\lub.\Delta \in \integer$
% then $\glb.\Delta = - \infty$. 
We define $\Inf.\Delta \sdef (\lub.\Delta = \infty)$, $\Fin.\Delta
\sdef \neg \Inf.\Delta$ and $\Empty.\Delta \sdef (\Delta =
\emptyset)$.  For a set $K$ and $i, j \in K$, we let $[i,j]_K \sdef
\{k : K \mid i \leq k \leq j\}$ denote the closed interval from $i$ to
$j$ containing elements from $K$.  One must often reason about two
\emph{adjoining} intervals, i.e., intervals that immediately precede
or follow a given interval. We say $\Delta$ adjoins $\Delta'$ iff
$\Delta \adjoins \Delta'$, where 
$$\begin{array}[t]{rcl}
  \Delta \adjoins \Delta' & \sdef &
  % \neg \Empty.\Delta\land \neg \Empty.\Delta' \imp
    \begin{array}[t]{@{}l}
      (\all t : \Delta, t' : \Delta' \st t < t') \land % (\Delta \cap \Delta' =
      % \emptyset) \land \\
      (\Delta \cup \Delta' \in Intv)
  \end{array}
\end{array}$$
\noindent
Note that adjoining intervals $\Delta$ and $\Delta'$ must be disjoint,
and by conjunct $\Delta \cup \Delta' \in Intv$, the union of $\Delta$
and $\Delta'$ must be contiguous. Note that both $\Delta \adjoins
\emptyset$ and $\emptyset \adjoins \Delta$ hold trivially for any
interval $\Delta$.

A \emph{stream} of behaviours over $VA \subseteq Var \cup Addr$ is
given by a total function of type $Stream_{VA} \sdef Time \fun
State_{VA}$, which maps each time to a state over $VA$. To reason
about specific portions of a stream, we use \emph{interval
  predicates}, which have type $IntvPred_{VA} \sdef Intv \fun
Stream_{VA} \fun \bool$. Note that because a stream encodes the
behaviour over all time, interval predicates may be used to refer to
the states outside a given interval.
% e.g., if $g_1$ and $g_2$ are interval predicates, $\Delta$ is an
% interval and $s$ is a stream, we have $(g_1 \land g_2).\Delta.s =
% (g_1.\Delta.s \land g_2.\Delta.s)$.
Like Interval Temporal Logic
\cite{Mos00}, we may define a number of operators on interval
predicates. For example, if $g \in IntvPred_{VA}$,
$\Delta\in Intv$ and $s \in Stream_{VA}$, we define:
$$
\begin{array}[t]{r@{\qquad\qquad}l}
  (\Box
  g).\Delta.s  \sdef   \all \Delta' : Intv \st \Delta' \subseteq \Delta \imp
  g.\Delta'.s
  &
  (\prev g).\Delta.s  \sdef  \exists
  \Delta' \st \Delta' \adjoins \Delta \land g.\Delta'.s
\end{array}
$$
\noindent
We assume pointwise lifting of operators on stream and interval
predicates in the normal manner, define \emph{universal implication} $
g_1 \entails g_2 \sdef \all \Delta : Intv, s : Stream \st g_1.\Delta.s
\imp g_2.\Delta.s $ for interval predicates $g_1$ and $g_2$, and say
$g_1\equiv g_2$ holds iff both $g_1 \entails g_2$ and $g_2 \entails
g_1$ hold.

We
define two operators on interval predicates: \emph{chop}, which is
used to formalise sequential composition, and $\omega$-{\em
  iteration}, which is used to formalise a possibly infinite iteration
(e.g., a while loop). The \emph{chop} operator `;' is a basic operator
on two interval predicates \cite{Mos00,DDH12,DH12MPC}, where $(g_1 \ch
g_2).\Delta$ holds iff either interval $\Delta$ may be split into two
parts so that $g_1$ holds in the first and $g_2$ holds in the second,
or the least upper bound of $\Delta$ is $\infty$ and $g_1$ holds in
$\Delta$. The latter disjunct allows $g_1$ to formalise an execution
that does not terminate. Using chop, we define the possibly infinite
iteration (denoted $g^\omega$) of an interval predicate $g$ as the
greatest fixed point of $z = (g \ch z) \lor \Empty$, where the
interval predicates are ordered using `$\entails$' (see \cite{DHMS12}
for details).  We define
$$\begin{array}{rcl}
  (g_1 \ch g_2).\Delta.s & \sdef &
  \begin{array}[t]{@{}l@{}}
    \left(\begin{array}[c]{@{}l@{}}
        \exists \Delta_1, \Delta_2 : Intv  \st
        (\Delta = \Delta_1
        \cup \Delta_2) \land \qquad {}\\
        \hfill (\Delta_1 \adjoins \Delta_2) 
        \land g_1.\Delta_1.s \land g_2.\Delta_2.s
      \end{array}\right) \lor 
    (\Inf \land g_1).\Delta.s
  \end{array}
  \\
  g^\omega & \sdef & \nu z \st (g \ch z) \lor
  \Empty 
\end{array}$$
\noindent
In the definition of $g_1 \ch g_2$, interval $\Delta_1$ may be empty,
in which case $\Delta_2 = \Delta$, and similarly $\Delta_2$ may empty,
in which case $\Delta_1 = \Delta$. Hence, both $(\Empty \ch g) \equiv
g$ and $g \equiv (g \ch \Empty)$ trivially hold. An iteration
$g^\omega$ of $g$ may iterate $g$ a finite (including zero) number of
times, but also allows an infinite number of iterations \cite{DHMS12}.

\paragraph{Permissions and interference.}  
To model true concurrency, the behaviour of the parallel composition
between two processes in an interval $\Delta$ is modelled by the
conjunction of the behaviours of both processes executing within
$\Delta$. Because this potentially allows conflicting accesses to
shared variables, we incorporate fractional permissions into our
framework \cite{Boy03,DDH12}.  We assume the existence of a {\em
  permission variable} in every state $\sigma \in State_{VA}$ of type
$VA \fun Proc \fun [0,1]_\rat$, where $VA \subseteq Var \cup Addr$ and
$\rat$ denotes the set of rationals.
% \begin{definition}
%   \label{def:read-write-perm}
A process $p \in Proc$ has {\em write-permission} to location $va\in
VA$ in $\sigma \in State_{VA}$ iff $\sigma.\Pi.va.p = 1$; has {\em
  read-permission} to $va$ in $\sigma$ iff $0 < \sigma.\Pi.va.p < 1$;
and has {\em no-permission} to access $va$ in $\sigma$ iff
$\sigma.\Pi.va.p = 0$.
% \end{definition}
% Note that unlike some approaches (e.g., \cite{BCOP05,RG09}), where $p$
% has read access to $v$ at time $t$ if $\Pi_p.v.t \neq 1$, we
% epplicitly model $\Pi_p.v.t = 0$ to mean ``no access''. This level of
% control is necessary because reads and writes to a variable may occur
% in a truly concurrent
% manner. % Before performing a write to a location, the
% % process performing the write must ensure that all read permissions to
% % the location have been returned \cite{Boy03}.

We define $\mcR.va.p.\sigma \sdef (0 < \sigma.\Pi.va.p < 1)$ and
$\mcW.va.p.\sigma \sdef (\sigma.\Pi.va.p = 1)$ and $\mcD.va.p.\sigma
\sdef (\sigma.\Pi.va.p = 0)$ to be state predicates on permissions.
In the context of a stream $s$, for any time $t \in \integer$, process
$p$ may only write to and read from $va$ in the transition step from
$s.(t-1)$ to $s.t$ if $\mcW.va.p.(s.t)$ and $\mcR.va.p.(s.t)$ hold,
respectively. % Otherwise, i.e., if $\mcD_p.v.(s.t)$ holds, process $p$
% may not access $v$ in the transition from $s.(t-1)$ to $s.t$.
Thus, $\mcW.va.p.(s.t)$ does not give $p$ permission to write to $va$
in the transition from $s.t$ to $s.(t+1)$ (and similarly $\mcR.va.p$).
For example, to state that process $p$ updates variable $v$ to value
$k$ at time $t$ of stream $s$, the effect of the update should imply
$((v = k) \land \mcW.v.p).(s.t)$.

One may introduce healthiness conditions on streams that formalise our
assumptions on the underlying hardware. We assume that at most one
process has write permission to a location $va$ at any time, which is
guaranteed by ensuring the sum of the permissions of the processes on
$va$ at all times is at most $1$, i.e., $$\all s: Stream, t: Time \st
((\displaystyle\Sigma_{p \in Proc} \Pi.va.p) \leq 1).(s.t)$$ Other
conditions may be introduced to model further restrictions as required
\cite{DDH12}.

Fractional permissions may also be used to characterise interference
within a process $p$. For a set of variables, we define $\mcI.VA.p 
\sdef  \exists v : VA \st \exists q : Proc \bs p \st \mcW.v.q$. % :
% \begin{eqnarray*}
%   \mcI.VA.p   & \sdef & \exists v : VA \st \exists q : Proc \bs p \st \mcW.v.q
% \end{eqnarray*}
Such notions are particularly useful because we aim to develop
rely/guarantee-style reasoning, where we use rely conditions to
characterise the behaviour of the environment. One may introduce rely
conditions that refer to $\mcI.VA.p$ to characterise the interference
on $VA$ by the environment of $p$.

\section{Evaluating state predicates over intervals}
\label{sec:eval-state-pred}

The set of times within an interval corresponds to a set of states
with respect to a given stream. Hence, if one assumes that expression
evaluation is non-atomic (i.e., takes time), one must consider
evaluation with respect to a set of states, as opposed to a single
state. It turns out that there are a number of possible ways in which
such an evaluation can take place, with varying degrees of
non-determinism \cite{HBDJ13}. In this paper, we consider \emph{actual
  states evaluation}, which evaluates an expression with respect to
the set of actual states that occur within an interval and
\emph{apparent states evaluation}, which considers the set of states
apparent to a given process.

Actual states evaluation allow one to reason about the true state of a
system, and evaluates an expression instantaneously at a single point
in time. However, a process executing with fine-grained atomicity can
only read a single variable at a time, and hence, will seldom be able
to view an actual state because interference may occur between two
successive reads. For example, a process $p$ evaluating $ecl_3$ (the
expression at $cl_3$) cannot read both $n1_p\mapsto mrk$ and
$n1_p\mapsto val$ in a single atomic step, and hence, may obtain a
value for $ecl_3$ that is different from any actual value of $ecl_3$
because interference may occur between reads to $n1_p\mapsto mrk$ and
$n1_p\mapsto val$. Therefore, we define an apparent states evaluator
that models fine-grained expression evaluation over intervals. Our
definition of apparent states evaluation does not fix the order in
which $n1_p\mapsto mrk$ and $n1_p\mapsto val$ are read. We see this as
advantageous over frameworks that must make the atomicity explicit
(e.g., \cite{VHHS06,CGLM06,DSW11}), which require an ordering to be
chosen, even if an evaluation order is not specified by the
corresponding implementation (e.g., \cite{HHLMSS07}). In
\cite{VHHS06,CGLM06,DSW11}, if the order of evaluation is modified,
the linearisability proof must be redone, whereas our proof is more
general because it shows that any order of evaluation is valid.

  % The set of apparent states may differ from the set of
% actual states.

% There are different methods for evaluating expressions with respect to
% a given interval and stream \cite{DH12MPC,DDH12,HBDJ11}. The
% differences arise from the underlying atomicity assumptions of the
% systems under consideration. Given that $k$ is an address, $va$ is a
% location, $he$ is an address-valued expression, $e$ is an expression,
% $\uop$ is a unary operator, $\bop$ is a binary operator and $f$ is a
% field, we define an evaluation function $eval$ over a state $\sigma$
% as follows.
% $$

% \paragraph{Evaluation at the ends of an interval.} 
\smallskip

\noindent{\bf Evaluation over actual states.}
% Most implementations only guarantee that at most one global variable
% can be read in a single atomic step. Thus, in the presence of possibly
% interfering processes and fine-grained atomicity, an evaluation model
% that assumes that a state predicate containing multiple variables can
% be evaluated in a single state may not be implementable
% \cite{CJ07,JP11}. That is, although an implementation evaluates a
% state predicate using a number of fine-grained atomic steps, this
% reality is not reflected in a model that assumes coarse-grained
% atomicity.  Hence, 
% The framework we develop assumes fine-grained interleaving and assumes
% that evaluation of a state predicate occurs over an interval of
% time. 
% Due to interference from other processes, the values of the
% variables (store) and addresses (heap) may be changing during this
% interval. Hence, we consider methods for non-determinis\-ti\-cally
% evaluating state predicates over an observation interval
% \cite{HBDJ11,DDH12}
% the inerTo this end, we consider the sets of states and sets of
% apparent states evaluators, which we introduce using Examples
% \ref{ex:setofstates} and \ref{ex:apparent}, respectively.
To formalise evaluators over actual states, for an interval $\Delta$
and stream $s \in Stream_{VA}$, we define $states.\Delta.s \sdef
\{\sigma : State_{VA} \mid \exists t : \Delta \st \sigma = s.t
\}$. Two useful operators for a sets of actual states of a state
predicate $c$ are $\Sometime c$ and $\Always c$, which specify that
$c$ holds in \emph{some} and \emph{all} actual state of the given
stream within the given interval, respectively. 
$$\begin{array}[t]{rcl}
  (\Sometime c).\Delta.s \sdef  \exists \sigma : states.\Delta.s \st
  c.\sigma 
  & \qquad & 
  (\Always c).\Delta.s \sdef \all \sigma : states.\Delta.s \st
  c.\sigma 
\end{array}$$

\begin{example} \label{ex:uv} Suppose $v$ is a variable, $fa$ and $fb$
  are fields, and $s$ is a stream such that the expression $(v \mapsto
  fa, v \mapsto fb)$ always evaluates to $(0, 0)$, $(1, 0)$ and $(1,
  1)$ within intervals $[1, 4]_\nat$, $[5,10]_\nat$ and
  $[11,16]_\nat$, respectively, i.e., for example $\Always ((v \mapsto
  fa, v \mapsto fb) = (0, 0)). [1, 4]_\nat.s$. Thus, both $\Always ((v
  \mapsto fa) \geq (v \mapsto fb)).[1,16]_\nat.s$ and $\Sometime ((v
  \mapsto fa) > (v \mapsto fb)).[1,16]_\nat.s$ may be deduced.
\end{example}

Using $\Always$, we define $\ola{c}$ and $\ora{c}$, which hold iff $c$
holds at the beginning and end of the given interval, respectively. 
$$
\begin{array}[t]{r@{\qquad\qquad}l}
  \ola{c}.\Delta.s
  \sdef (\Always c \land \neg \Empty) \ch true
  &
  \ora{c}.\Delta.s \sdef true \ch (\Always c \land \neg \Empty)
\end{array}
$$
Operators $\Always$ and $\Sometime$ cannot accurately model
fine-grained interleaving in which processes are able to access at
most one location in a single atomic step. However, both $\Always$ and
$\Sometime$ are useful for modelling the actual behaviour of the
system as well as the behaviour of the coarse-grained abstractions
that we develop. % We may use $\Always$ to define \emph{invariance} of a
% state predicate $c$ using $inv.c \sdef  \prev \ora{c} \imp \Always c$.% follows:
% % \begin{eqnarray*}
% \end{eqnarray*}
We may use $\Always$ to define \emph{stability} of a
variable $v$, and \emph{invariance} of a state predicate $c$ as
follows:
$$\begin{array}[t]{rcl}
  stable.v  \sdef        \exists k \st \prev \ora{(va = k)} \land \Always (va
      = k) & \qquad\qquad &
    %   \left\{
    % \begin{array}[c]{@{}l@{\qquad}l@{}}
    %   \exists k \st \prev \ora{(va = k)} \land \Always (va
    %   = k) & \textrm{if $va \in Var$} \\
    %   \exists k \st \prev \ora{(\deref va = k)} \land
    %   \Always (\deref va
    %   = k) & \textrm{if $va \in Addr$} 
    % \end{array}\right.
  inv.c \sdef  \prev \ora{c} \imp \Always c
\end{array}$$
\noindent
Such definitions of stability and invariance are necessary because
adjoining intervals are assumed to be disjoint, i.e., do not share a
point of overlap. Therefore, one must refer to the values at the end
of some immediately preceding interval. \smallskip

\noindent{\bf Evaluation over states apparent to a process.}
Assuming the same setup as \refex{ex:uv}, if $p$ is only able to
access at most one location at a time, evaluating $(v \mapsto fa) < (v
\mapsto fb)$ using the states \emph{apparent} to process $p$ over the
interval $[1,16]_\nat$ may result in $true$, e.g., if the value at $v
\cdota fa$ is read within interval $[1,4]_\nat$ and the value at $v
\cdota fb$ read within $[11,16]_\nat$. % (cf. \cite{DDH12,HBDJ11}).

Reasoning about the apparent states with respect to a process $p$
using function $apparent$ is not always adequate because it is not
enough for an apparent state to exist; process $p$ must also be able
to read the relevant variables in this apparent state. Typically, it
is not necessary for a process to be able to read all of the state
variables to determine the apparent value of a given state
predicate. In fact, in the presence of local variables (of other
processes), it will be impossible for $p$ to read the value of each
variable. Hence, we define a
function % s $substates_{p,W}.\Delta.s$ and
$apparent_{p,W}$, where $W \subseteq Var \cup Addr$ is the set of
locations whose values process $p$ needs to determine to evaluate the
given state predicate.
$$\begin{array}[t]{rcl}
  % substates_{p,W}.\Delta.s & \sdef & \{\sigma : State_W \mid  \epists
  % t : \Delta \st \sigma = s.t \land (\all v : V \st \mcR_p.v.(s.t))\}
  % \\simpsi
  apparent_{p,W}.\Delta.s & \sdef &  \{\sigma : State_W \mid 
  \forall va : W \st \exists t : \Delta \st (\sigma.va = s.t.va) \land \mcR.va.p.(s.t)
  % \mcR_p.v.t
  \} 
\end{array}$$
\noindent
Using this function, we are able to determine whether state predicates
definitely and possibly hold with respect the apparent states of a
process. For a state predicate $c$, interval $\Delta$, stream $s$ and
state $\sigma$, we let $accessed.c.\sigma$ denote the smallest set of
locations (variables and addresses) that must be accessed in order to
evaluate $c$ in state $\sigma$ and define $locs.c.\Delta.s \sdef
\bigcup_{t \in \Delta} accessed.c.(s.t)$. For a process $p$, this is
used to define $(\Def_p\ c).\Delta.s$, which states that $c$ holds in
all states apparent to $p$ in $s$ within $\Delta$. (Similarly
$(\Pos_p\ c).\Delta.s$.)
$$
\begin{array}{rcl}
    (\Def_p\ c).\Delta.s &   \sdef &   \klet W = locs.c.\Delta.s \kin  \all \sigma :
    apparent_{p,W}.\Delta.s \st  c.\sigma \\
    (\Pos_p\ c).\Delta.s &  \sdef &  \klet W = locs.c.\Delta.s \kin  \exists \sigma :
    apparent_{p,W}.\Delta.s \st  c.\sigma
\end{array}$$
\noindent 
Continuing \refex{ex:uv}, if $c \sdef ((v \mapsto fa) \geq (v \mapsto
fb))$, we have $(\neg \Defp c).[1,16]_\nat.s$ holds, i.e., $(\Posp
\neg c).[1,16]_\nat.s$ even though $(\Always c).[1,16]_\nat.s$ holds
(cf. \cite{DDH12,HBDJ13}).
% \end{example}
One may establish a number of properties on $\Always$, $\Sometime$,
$\Def$ and $\Pos$ \cite{HBDJ13}, for example
% \begin{eqnarray}
%   \label{eq:16}
$\Pos_p (c \land d) \entails \Pos_p c \land \Pos_p d$ holds.
% \end{eqnarray}
The following lemma relates apparent and states evaluation.
\begin{lemma}
  \label{lem:stable}
  For any process $p$, variable $v$, field $f$ and constant $k$,

  $stable.v \land \Pos_p ((v \mapsto f) = k) \imp \Sometime((v \mapsto f) =
  k)$
\end{lemma}

\section{Behaviours and refinement}
\label{sec:behaviour-refinement-1}

% \subsection{Semantics}
% \label{sec:semantics}

The \emph{behaviour} of a command $C$ executed by a non-empty set of
processes $P$ in a context $Z \subseteq Var$ is given by interval
predicate $beh_{P,Z}.C$, which is defined inductively in
\reffig{fig:beh-def}. We use $beh_{p,Z}$ to denote $beh_{\{p\},Z}$ and
assume the existence of a program counter variable $pc_p$ for each
process $p$. We define shorthand $\FinIdle \sdef \Enf \Fin \Idle$ and
$\InfIdle \sdef \Enf \Inf \Idle $ to denote finite and infinite
idling, respectively and use the interval predicates below to
formalise the semantics of the commands in
\reffig{fig:beh-def}.
$$\begin{array}[t]{rcl}
  \Eval_{p,Z}.c & \sdef & 
  \Posp c \land 
  beh_{p,Z}.\Idle
  \\
  % \mathsf{update}_{p,Z}(va,k) & \sdef &
  % \Always ((va = k) \land \mcW.va.p \land beh_{p, Z\bs\{va\}}.\FinIdle) \land 
  % \neg \Empty
  \Update_{p,Z}(va,k) & \sdef &  
 \!\left\{\begin{array}[c]{@{}l@{\:\:\:}l@{}}
      beh_{p, Z \bs\{va\}}.\Idle
    \land \neg \Empty \land \Always(va = k \land \mcW_p.va) &
    \textrm{if $va \in Var$} \\
    beh_{p, Z \bs\{va\}}.\Idle
    \land \neg \Empty \land \Always((\deref\! va) = k \land \mcW_p.va) &
    \textrm{if $va \in Addr$} 
  \end{array} \right.
\end{array}
$$
\noindent
To enable compositional reasoning, for interval predicates $r$ and
$g$, and command $C$, we introduce two additional constructs
$\Rely{r}{C}$ and $\Enf{g}{C}$, which denote a command $C$ with a
\emph{rely condition} $r$ and an \emph{enforced condition} $g$,
respectively \cite{DDH12}.

\begin{figure}[t]
  \figrule \centering \small

  $
  \begin{array}{@{}rcl@{}}
    \begin{array}[t]{rcl}
      % beh_{p,Z}.\Chaos & \sdef & true
      % \\
      beh_{p,Z}.\Idle & \sdef & \all va : Z \st \Always \neg \mcW.va.p
      \\
      beh_{p,Z}.{[}c{]} & \sdef &   \Posp c \land 
  beh_{p,Z}.\Idle
% \mathsf{apparently}_{p,Z}.c
      \\
      beh_{p,Z}.\AG{c} & \sdef & % \mathsf{eventually}_{p,Z}.c 
      \Sometime c \land beh_{p,Z}.\Idle
      \\
      beh_{P,Z}.C^\omega & \sdef & (beh_{P,Z}.C)^\omega 
      \\
      beh_{p, Z}.(l : C) & \sdef & \Always(pc_p = l) \land beh_{p,Z}.C
    \end{array}
    &\quad & 
  \begin{array}[t]{rcl}
    beh_{P,Z}.(C_1 \ch C_2) & \sdef &  beh_{P,Z}.C_1 \ch
    beh_{P,Z}.C_2
     \\
    beh_{P,Z}.(C_1 \sqcap C_2) & \sdef &  beh_{P,Z}.C_1 \lor
    beh_{P,Z}.C_2
     \\
    beh_{P,Z}.(\Rely{r}{C}) &  \sdef &  r \imp beh_{P,Z}.C 
    \\
    beh_{P,Z}.(\Enf{g}{C}) & \sdef &  g \land beh_{P,Z}.C 
  \end{array}
  \\
  \multicolumn{3}{l}{
    \begin{array}[t]{rcl}
      beh_{p,Z}.(vae \asgn e) & \sdef &
      \left\{
        \begin{array}[c]{@{}l@{\quad\:\:}l}
          \exists k \st \Eval_{p,Z}.(e=k) % beh_{p,Z}.[e = k]
          \ch 
          \mathsf{update}_{p,Z}(v,k)      
          & \textrm{if $vae \in Var$}
          \\
          \exists k, a\st 
          \Eval_{p,Z}.(vae = a \land e = k)
          \ch \mathsf{update}_{p,Z}(a,k)      
          & 
          \textrm{otherwise}
        \end{array}
      \right.
       \\
      beh_{P,Z}.(\Par_{p:P}\ C_p) & \sdef & \\
      \multicolumn{3}{l}{
        \qquad \begin{array}[t]{@{}l@{}}
      \left\{
        \begin{array}[c]{@{}l@{\qquad}l}
          true & \textrm{if $P = \emptyset$}  \\
          beh_{p,Z}.C_p & \textrm{if $P = \{p\}$}  \\
          \begin{array}[t]{@{}l@{}}
          \exists P_1, 
          P_2, S_1, S_2 \st 
          \begin{array}[t]{@{}l@{}}
            (P_1 \cup P_2 = P) \land  
            (P_1 \cap P_2 = \emptyset) \land P_1 \neq \emptyset \land
            P_2 \neq \emptyset \land \\
            S_1 \in \{\FinIdle, \InfIdle\} \land S_2 \in
            \{\FinIdle, \InfIdle\} \land  \\
            (S_1 = 
            \InfIdle \imp S_2 \neq \InfIdle)  \land
            \\
            beh_{P_1,Z}.((\Par_{p:P_1} C_p) \ch S_1)
            \land 
            beh_{P_2,Z}. ((\Par_{p:P_2} C_p) \ch S_2) 
          \end{array}
          \end{array}
          & \textrm{otherwise} 
        \end{array}\right. 
    \end{array}
    }
      \\
      beh_{P, Z}.\Context{Y}{C} & \sdef &
      (Z \cap Y = \emptyset) \land beh_{P,Z
        \cup Y}.C % \land fresh.Y 
    \end{array}
  }
  \end{array}$

  \figrule
  \caption{Formalisation of behaviour function}
  \label{fig:beh-def}
\end{figure}

We say that a concrete command $C$ is a refinement of an abstract
command $A$ iff every possible behaviour of $C$ is a possible
behaviour of $A$. Command $C$ may use additional variables to those in
$A$, hence, we define refinement in terms of sets of variables
corresponding to the contexts of $A$ and $C$. In particular, we say
$A$ with context $Y$ is \emph{refined} by $C$ with context $Z$ with
respect to a set of processes $P$ (denoted $A \sref_P^{Y,Z} C$) iff
$beh_{P,Z}.C \entails beh_{P,Y}.A$ holds. Thus, any behaviour of the
concrete command $C$ is a possible behaviour of the abstract command
$A$. This is akin to operation refinement \cite{deRoever98}, however,
our definition is with respect to the intervals over which the
commands execute, as opposed to their pre/post states. We write $A
\sref_P^{Z} C$ for $A \sref_P^{Z, Z} C$, write $A \sref_P C$ for $A
\sref_P^\emptyset C$, write $A \srefeq_P^{Z} C$ iff both $A
\sref_P^{Z} C$ and $C \sref_P^{Z} A$, and write $A \sref^{Y,Z}_p C$
for $A \sref_{\{p\}}^{Y,Z} C$. There are numerous theorems and lemmas
for behaviour refinement \cite{DDH12,DD12}. We present a selection of
results that are used to verify correctness of the lazy set. The
following results may be proved using monotonicity of the
corresponding interval predicate operators.

% \begin{lemma}
%   \label{lem:command-ref}
%   Suppose $p$ is a process and $P$ is a non-empty set of processes,
%   $A$, $A_1$, $A_2$, $C$, $C_1$ and $C_2$ are commands and $Y, Z
%   \subseteq Var$ such that $A \sref_P^{Y,Z} C$, $A_1 \sref_P^{Y,Z}
%   C_1$ and $A_2 \sref_P^{Y,Z} C_2$ hold. Then, provided $c \imp b$ % and
%   % $\all k : Val \st \Pos_p (e' = k) \entails \Pos_p (e = k)$ hold,
%   each of the following refinements hold.\smallskip
  
%   \quad$
%   \begin{array}{@{}l@{\qquad\qquad}l@{\qquad\qquad}l@{}}
%     \begin{array}[t]{rcl}
%       [b] & \sref_p^{Y,Z} & [c] \smallskip \\
%       \AG{b} & \sref_p^{Y,Z} & \AG{c} \smallskip \\
%       % v \asgn e & \sref_p^{Y,Z} & v \asgn e' \smallskip \\
%       % v \hasgn e & \sref_p^{Y,Z} & v \hasgn e' 
%     \end{array}
%     & 
%     \begin{array}[t]{rcl}
%       A_1 \ch A_2 & \sref_P^{Y,Z} & C_1 \ch C_2 \smallskip \\
%       A_1 \sqcap A_2 & \sref_P^{Y,Z} & C_1 \sqcap C_2 \smallskip \\
%     \end{array}
%     & 
%     \begin{array}[t]{rcl}
%       A^\omega & \sref_P^{Y,Z} & C^\omega \smallskip \\
%       \rely r \st A & \sref^{Y,Z}_P & \rely r \st C
%     \end{array}
%   \end{array}
%   $
% \end{lemma}
The next lemma states that an assignment of state predicate $c$ to a
variable $v$ may be decomposed to a guard $[c]$ followed by an
assignment of $true$ to $v$ and a guard $[\neg c]$ followed by an
assignment of $false$ to $v$.
\begin{lemma}
  \label{lem:bool-exp}
  For a state predicate $c$, variable $v$, process $p$, and
  $Z\subseteq Var\cup Addr$, we have

  $v \asgn c \sref_p^{Z}
  ([c] \ch v \asgn true) \sqcap ([\neg c] \ch v \asgn false)$.
\end{lemma}
Note that a property like \reflem{lem:bool-exp} is difficult to
formalise in interleaved frameworks such as action systems
\cite{BvW99} because interference may occur between guard evaluation
and assignment to $v$ at the concrete level, which is not possible in
the abstract. The lemma below allows one to move the frame of a
command into the refinement relation.
\begin{lemma}
  \label{lem:add-context}
  Suppose $A$ and $C$ are commands, $P \subseteq Proc$, $W,X \subseteq
  Var$ and $Y,Z \subseteq Var \cup Addr$ such that $W \subseteq (X
  \cup Z)$ and $W \cap Y = \emptyset = X \cap Z$. If $A \sref^{W\cup
    Y, X \cup Z }_P C$, then $\Context{W}{A} \sref_P^{Y,Z}
  \Context{X}{C}$.
\end{lemma}

The following theorem allows one to turn a rely condition at the
abstract level to an enforced condition at the concrete level,
establishing a Galois connection between rely and enforced conditions
\cite{DDH12}.
\begin{theorem}
  \label{thm:rely-enf-gc}
  $(\Rely{r}{A}) \sref_P^{Y,Z} C\:\: \iff\:\: A \sref_P^{Y,Z} (\Enf{r}{C})$
\end{theorem}
% There are several rules for compositionally refining commands
% \cite{DDGH12}. We present a number of proofs that are useful for
% verifying Heller et al's algorithm. 
When modelling a lock-free algorithm \cite{CGLM06,DSW11,VHHS06}, one
assumes that each process repeatedly executes operations of the data
structure, and hence the processes of the system only differ in terms
of the process ids. For such programs, a proof of the parallel
composition may be decomposed using the following theorem \cite{DD12}.
\begin{theorem}
  \label{thm:decompose}
  If $p \in Proc$, $Y,Z \subseteq Var \cup Addr$, and
  $A(p) % \Context{W}{A(p)^\omega}
  $ and $C(p)% \Context{X}{C(p)^\omega}
  $ are commands parameterised by $p$, then $
  (\Rely{g}{\Par_{p:P}A(p)}) \sref_P^{Y,Z} (\Par_{p:P}C(p))$ holds if
  for some interval predicate $r$ and some $p \in P$ and $Q \sdef P
  \bs \{p\}$ both of the following hold.
  \begin{eqnarray}
    \label{eq:17}
    \Rely{g \land r}{A(p)} & \sref_{p}^{Y,Z} &
    C(p)% \Rely{r \land r}{A(p)} & \sref_{p}^{Y,Z} & C(p)
    \\
    \label{eq:18}
    g \land beh_{Q,Z}.(\Par_{q:Q}C(q)) &  \entails &
    r
  \end{eqnarray}
\end{theorem}

\begin{figure}[!t]
  \centering
  \figrule\small

    $\begin{array}[t]{@{}rcl@{}}
      \varphi^{k+1}.ua.\sigma &\sdef& 
      \begin{array}[t]{@{}l@{}}
      \kif (k = 0) \kthen 
      ua       \kelse\ 
      eval.((\varphi^k.ua.\sigma) \mapsto nxt).\sigma
    \end{array}
      \\
      \reachable.ua.vb.\sigma &\sdef&  \exists k : \nat \st
      \varphi^k.ua.\sigma = vb
      \\
      setAddr.\sigma & \sdef & \left\{a :
      Addr 
      \begin{array}[c]{@{~}|@{~}l@{}}
        \reachable.Head.a.\sigma  \land 
        \neg eval.(a\mapsto mrk).\sigma
      \end{array}
    \right\} 
      \\
      absSet.\sigma & \sdef & \left\{v :
      Val 
      \begin{array}[c]{@{~}|@{~}l@{}}
        \exists a : setAddr.\sigma \st 
        \hfill v = eval.(a \mapsto
        val).\sigma
      \end{array}
    \right\} 
      \smallskip\\
      \mathsf{CGCon}(p,x) & \sdef &
      \begin{array}[t]{@{}rl@{}}
        (\AG{x \in absSet} \ch res_p \asgn true )
        \sqcap  {}
        (\AG{x \notin absSet} \ch res_p \asgn false)
      \end{array}
    \end{array}$
    \smallskip
  
    $\begin{array}[t]{@{}rcl@{}} 
      \mathsf{CGS}(p) & \sdef &
      \Context{res_p}{\left( \bigsqcap_{x : \integer} \left(
            \begin{array}[c]{@{}ll@{}}
              
              \Context{
                \begin{array}[c]{@{}r@{}}
                  n1_p, n2_p, n3_p
                \end{array}
              }{
                \mathsf{Add}(p,x) 
                \sqcap
                \mathsf{Remove}(p,x)}
              {}\sqcap   \mathsf{CGCon}(p,x)
            \end{array}
          \right) \right)^{\omega}}
      \smallskip \\
      \mathsf{CGSet}(P) & \sdef & \Context{Head,Tail}{\rely
        \ola{HTInit} \st \Par_{p:P}\ 
        \mathsf{CGS}(p)}
    \end{array}$
    \figrule
  \caption{A coarse-grained abstraction of {\tt contains}}
  \label{fig:labs}
\end{figure}

\section{Verification of the lazy set}
\label{sec:verif-lazy-set}

As already mentioned, we focus on a proof {\tt contains}, which
highlights the advantages of interval-based reasoning over frameworks
that only reason about the pre/post states.\footnote{A verification of
  the {\tt add} and {\tt remove} operations are presented in Appendix
  A.}  Verification of linearisability of \texttt{contains} is known
to be difficult using frameworks that only consider the pre/post
states \cite{Vaf10,VHHS06,CGLM06,DSW11}. A coarse-grained abstraction
of $\mathsf{Set}(P)$ in \reffig{fig:formal-lazyset} is given by
$\mathsf{CGSet}(P)$ in \reffig{fig:labs}, where the $\mathsf{Add}$ and
$\mathsf{Remove}$ operations are unmodified, but $\mathsf{Contains}$
is replaced by $\mathsf{CGCon}$, which tests to see if $x$ is in the
set using an atomic (coarse-grained) guard, then updates the return
value to $true$ or $false$ depending on the outcome of the test.

State predicates $reachable$, $setAddr$ and $absSet$, which are used
our refinement proof, are defined in \reffig{fig:labs}. A location
$vb$ is \emph{reachable} from $ua$ in state $\sigma$ iff
$\reachable.ua.vb.\sigma$ holds, hence, for example,
$\reachable.Head.n.\sigma$ holds iff it is possible to traverse the
list starting from $Head$ and reach node $n$ in $\sigma$. The abstract
set of node addresses corresponding to each list data structure in
$\sigma$ is given by $setAddr$ and the set of values of these nodes is
given by $absSet.\sigma$. Although $null$ is always reachable from
$Head$, $setAddr$ will not contain $null$ because $null \notin Addr$.

\begin{figure}[t]
  \centering
  \scalebox{0.65}{\input{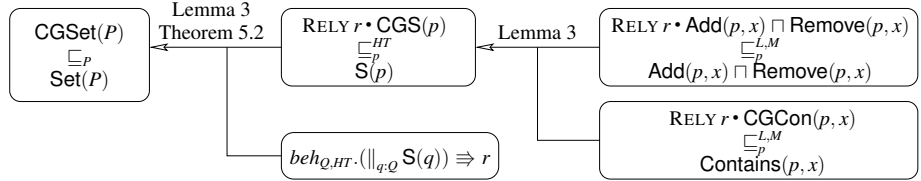}}
  \caption{Proof decomposition for the lazy set verification}
  \label{fig:refsteps}
\end{figure}

An overview of the proof decomposition is given in
\reffig{fig:refsteps}. 
% where the parameters to the $\sref$ relation
% have been omitted.
To prove that $\mathsf{Set}(P)$ refines $\mathsf{CGSet}(P)$, using
\refthm{thm:decompose} we show that $\mathsf{S}(p)$ refines
$\mathsf{CGS}(p)$ for a single process $p \in P$ under a yet to be
determined rely condition $r$ (condition \refeq{eq:17}), provided that
the behaviour of the rest of the program implies the $r$ that is
derived (condition \refeq{eq:18}). Then, using monotonicity of $\sref$
% Lemmas \ref{lem:command-ref} and
and \reflem{lem:add-context}, we further decompose the proof that
$\mathsf{S}(p)$ refines $\mathsf{CGS}(p)$ to the level of each
operation. The proofs for ${\sf Add}$ and ${\sf Remove}$ are trivial
because they are unmodified in $\mathsf{CGS}(p)$. To prove ${\sf
  Contains}$, we use \reflem{lem:bool-exp} to perform case analysis on
executions that return $true$ and $false$. The refinement proof is
hence localised as much as possible.  Furthermore, the structure of
$r$ is elucidated as part of the correctness proof.

% and $setAddrFields.\sigma$
% denotes the reachable field addresses, without the lock field. This is
% used by ${\sf LLocate}$, which in addition to satisfying the
% postcondition given by $LocPost$, must not modify the abstract set
% corresponding to the linked list.

% Operations $\mathsf{LAdd}$ and ${\sf LRemove}$ operations must ensure
% that there is no interference to fields {\tt val}, {\tt nxt} and {\tt
%   mrk} for the locked nodes. Hence, we define
% \begin{eqnarray*}
%   LAddr.V.\sigma & \sdef & \textstyle\bigcup_{v:V}
%   setFields.(\sigma.v)
%   \\
%   LMV.v.\sigma & \sdef & \{eval.((v \mapsto nxt)\cdota val).\sigma, eval.((v \mapsto nxt)\cdota mrk).\sigma\}
% \end{eqnarray*}
% \noindent
% Within command $\mathsf{LAdd}$, non-interference on variables $n1_p$,
% $n2_p$ and $n3_p$ (in the store) is guaranteed because they are local
% variables of process $p$ and non-interference on the addresses of
% $n1_p$, $n2_p$ and $n3_p$ (in the heap) is guaranteed by the enforced
% property.

We are required to prove $\mathsf{CGSet}(P) \sref_P \mathsf{Set}(P)$
for an arbitrarily chosen set of processes $P \subseteq Proc$. Using
\reflem{lem:add-context}, we transfer the context $HT \sdef Addr \cup
\{Head, Tail\}$ of $\mathsf{CGSet}(P)$ and $\mathsf{Set}(P)$ into the
refinement relation. Then, using monotonicity of $\sref$ followed by
\refthm{thm:decompose}, we decompose the specifications into the
following proof obligations, where $p \in P$ and $Q \sdef P \bs \{p\}$
and the rely condition $r$ is yet to be developed.
$$\begin{array}[t]{rcl}
  S1 \sdef \Rely{r}{\mathsf{CGS}(p)}  \sref_{p}^{HT} \mathsf{S}(p) & \quad 
  \quad & 
  S2 \sdef beh_{Q,HT}.(\Par_{q : Q} \mathsf{S}(q))  \entails  r
\end{array}$$

\noindent \textbf{Proof of $S1$.}
%n1_p, n2_p, n2_p, 
Using \reflem{lem:add-context} to expand the context followed by
monotonicity of ${}^\omega$ and $\sqcap$, assuming $L \sdef HT \cup
\{res_p% , n1_p, n2_p, n3_p
\}$ and $M \sdef L \cup \{n1_p, n2_p, n3_p\}$, condition $S1$
decomposes as follows.
\begin{eqnarray}
  \label{eq:3}
  \Rely{r}{\Context{n1_p, n2_p, n3_p}{\left(
      \begin{array}[c]{@{}l@{}}
        \mathsf{Add}(p, x) {} \sqcap {} 
       \mathsf{Remove}(p,x)
    \end{array}
  \right)}}
  % \begin{array}[c]{@{}l@{}}
  %   \mathsf{Add}(p,x) 
  %   {} \sqcap {}  \mathsf{Remove}(p,x)
  % \end{array}
& \sref_{p}^{L,M} &
  \begin{array}[c]{@{}l@{}}
    \mathsf{Add}(p,x) 
    {} \sqcap {}  \mathsf{Remove}(p,x)
  \end{array}
  \\
  \label{eq:7}
  \Rely{r}{\mathsf{CGCon}(p, x)} & \sref_{p}^{L,M} & \mathsf{Contains}(p,x) 
\end{eqnarray}  
 % holds if $\mathsf{Add}$ ($\mathsf{Remove}$)
% guarantees non-interference on the addresses of $n1_p$, $n2_p$, and
% $n3_p$ ($n1_p$ and $n2_p$), which are straightforward due to the
% locks. Non-interference on $n2_p$ within $\mathsf{Add}$ is guaranteed
% because $n2_p$ is fresh.

% for the $\mathsf{Add}$ operation,
% non-interference on $n1_p$ and $n3_p$ is guaranteed by the locks and
% non-interference on $n2_p$ is guaranteed because $n2_p$ is a fresh
% node. For the $\mathsf{Remove}$ operation, non interference on $n1_p$
% and $n2_p$ are guaranteed by the locks and non-interference on $n3_p$
% holds by ensuring that $r$ implies:
% \begin{eqnarray}
%   \label{eq:14}
%   \begin{array}[t]{@{}l@{}}
%     \all p :Proc,  % v: Var, 
%     a : Addr \st \\
%     \quad \Always (
%     \begin{array}[c]{@{}l@{}}
%       \reachable.a \land ((a \mapsto lck) = p)
%       % \land ((a \mapsto nxt) = b) 
%     \end{array}\imp 
%       \hfill \neg \mcI.(setFields.(a \mapsto nxt)).p )
% \end{array}
% \end{eqnarray}
% That is, in all actual states, for any process $p$ and address $a$, if
% $a$ is reachable from $Head$ and the lock for $a$ is held by $p$,
% there cannot be any interference on the fields of $a \mapsto nxt$ with
% respect to $p$.

% \begin{brijesh}
%   Informal explanation of why this is correct.
% \end{brijesh}
Condition \refeq{eq:3} is trivial by \reflem{lem:add-context} and
reflexivity of $\sref_p^{M}$. To prove \refeq{eq:7}, must ensure that
if $res_p$ is assigned $true$, then there must have been an actual
state, say $\sigma$, in the interval preceding the assignment to
$res_p$ such that $x \in setVal.\sigma$. Similarly, if $res_p$ is
assigned $false$, there must have been an actual state $\sigma$ within
the interval of execution such that $x \notin setVal.\sigma$. Note
that in the proof, we use the states apparent to process $p$ to deduce
a property of an actual state of the system. Using
\reflem{lem:bool-exp}, ${\sf Contains}(p,x)$ is equivalent to the
following, where\smallskip

\noindent
\quad
$\begin{array}{rcl}
IN \sdef   \neg (n1_p\mapsto mrk)
\land ((n1_p\mapsto val) = x) & \quad & 
CL \sdef cl_1: (n1_p \asgn Head)
\ch cl_2: {\sf CLoop}(p,x)
\end{array}$\hfill\smallskip

\noindent
and split the label $cl_3$ into $clt_3$ and $clf_3$ --- the true and
false cases of $IN$.
$$  \begin{array}[c]{@{}ll@{}}
    % cl_1: n1_p \asgn Head \ch cl_2: {\sf CLoop}(p,x)
    CL
    \ch 
    \left ((clt_3: ([IN]
    \ch res_p \asgn true))
    \sqcap (clf_3:([\neg IN] \ch res_p \asgn false)) \right)
  \end{array}
$$
\noindent
We distribute $CL$ within the `$\sqcap$', use monotonicity to match
the abstract and concrete $true$ and $false$ branches, then use
monotonicity again to remove the assignments to $res_p$ from both
sides of the refinement. Thus, we are required to prove the following
properties.
\begin{eqnarray}
  \label{eq:4}
  \rely r \st \AG{x \in absSet}
  & 
  \sref_{P}^{L,M} & CL
  \ch clt_3:[IN]% \left[ 
  % \begin{array}[c]{@{}l@{}}
  %   \neg (n1_p\mapsto mrk) \land \\
  %   (n1_p\mapsto val) = x
  % \end{array}
  % \right]
  \\
  \label{eq:8}
  \rely r \st \AG{x \notin absSet}
  & 
  \sref_{P}^{L,M} & CL % cl_1: n1_p \asgn Head \ch cl_2: {\sf CLoop}(p,x)
  \ch clf_3:[\neg IN]% [n1_p\mapsto mrk \lor (n1_p\mapsto 
  % val) \neq x]  
\end{eqnarray}
Condition \refeq{eq:4} (i.e., the branch that assigns $res_p \asgn
true$) states that there must be an actual state $\sigma$ within the
interval in which $CL \ch clt_3: [IN]$ executes, such that $x \in
absSet.\sigma$ holds, which indicates that there is a point at which
the abstract set contains $x$. It may be the case that a process $q
\neq p$ has removed $x$ from the set by the time process $p$ returns
from the contains operation. In fact, $x$ may be added and removed
several times by concurrent add and remove operations before process
$p$ completes execution of $\mathsf{Contains}(p,x)$. However, this
does not affect linearisability of $\mathsf{Contains}(p,x)$ because a
state for which $x \in absSet$ holds has been found.  An execution of
$\mathsf{Contains}(p,x)$ that returns $true$ would only be incorrect
(not linearisable) if $true$ is returned and $\Always (x \notin
absSet)$ holds for the interval in which $CL \ch clt_3:[IN]$
executes. Similarly, we prove correctness of \refeq{eq:8} by showing
that is impossible for there to be an execution that returns $false$
if $\Always (x \in absSet)$ holds in the interval of execution.
% Hence, if $inSet(Head, n2_p, x)$ holds, then $p$ must have added $x$
% to the set $S$ using node $n2_p$. 
\smallskip

\noindent{\bf Proof of \refeq{eq:4}.} Using \refthm{thm:rely-enf-gc},
we transfer the rely condition $r$ to the right hand side as an
enforced property. We define state predicate $inSet(ua,x)$, which
states that $ua$ with value $x$ is in the abstract set, i.e.,
$inSet(ua, x) \sdef \reachable.Head.ua \land \neg (ua \mapsto mrk)
\land (ua \mapsto val = x)$.  We require
that $r$ implies the following.
\begin{eqnarray}
  \label{eq:5}
  % \Box (\Always (pc_p = cl_2) & \imp & 
  & inv.(\reachable.Head.n1_p \lor  (n1_p \!\mapsto\!
  mrk)) \\
  \label{eq:10}
  & \Box (\Always (pc_p = cl_3) \imp inv.(n1_p \mapsto mrk) \land \all k : Val \st inv.((n1_p
  \mapsto val) = k))
  % \label{eq:1}
  % \Box (\Always (pc_p = cl_3) & \imp & )  
\end{eqnarray}
% This proof is given in  Appendix A. 
The behaviour of the right hand side of \refeq{eq:4} then simplifies as follows. 
\begin{derivation*}
  \step{ beh_{p,M}.(\enf r \st % n1_p \asgn Head \ch cl: {\sf
      % CLoop}(p,x)
    CL
    \ch cl_3: [IN])}

  \trans{\equiv}{definition of $beh$}

  \step{r \land  (beh_{p,M}.CL \ch  beh_{p,M}.(cl_3: [IN]))}

  \trans{\entails}{definition of $beh$ and $n1_p$ is local to $p$}

  \step{r \land  (beh_{p,M}.CL \ch (stable.n1_p \land beh_{p,M}.(cl_3: [IN])))}

  \trans{\entails}{$\Posp (c \land d) \entails \Posp c \land
    \Posp d$ and \reflem{lem:stable}}
  % $\Posp ((v \mapsto f) = k) \imp \Sometime ((v \mapsto f) = k)$ }

  \step{r \land \left(
      \begin{array}[c]{@{}l@{}}
        beh_{p,M}.(cl_1: n1_p \asgn Head) \!\ch\! beh_{p,M}.(cl_2: {\sf
          CLoop}(p,x)) \!\ch \! \Sometime \neg 
        (n1_p\mapsto mrk) 
        \land \Sometime ((n1_p\mapsto val) = x)
      \end{array}
    \right)}

  \trans{\entails}{first chop: change context $n1_p \notin L$, second
    chop: assumption \refeq{eq:5}} 

  \step{r \land \left(
    \begin{array}[c]{@{}l@{}}
      beh_{p,L}.\Idle \ch \Always (\reachable.Head.n1_p \lor (n1_p \mapsto
      mrk)) \ch {}  (\Sometime \neg (n1_p\mapsto mrk) \land
      \Sometime ((n1_p\mapsto val) = x))
    \end{array}
  \right)}
\end{derivation*}
Focusing on just the second and third parts of the chop, because
$n1_p$ is not modified after $\mathsf{CLoop}$, and $r$ is assumed to
split, we obtain the following calculation.
\begin{derivation*}

  \step{\exists a : Addr \st r \land 
    \left(
      \begin{array}[c]{@{}l@{}}
        \left(\Always (\reachable.Head.n1_p \lor (n1_p \mapsto mrk)) \land \ora{n1_p = a}\right) \ch \qquad \\
        \hfill (\Always (n1_p = 
        a) \land \Sometime \neg (a \mapsto mrk) \land \Sometime (a \mapsto
        val) = x))
      \end{array}\right)
  }

  \trans{\entails}{$\Always c \entails \ora{c}$, then by assumption
    \refeq{eq:10}, disjunct $\ora{(a\mapsto 
      mrk)}$ in LHS of chop }

  % \step{\exists a : Addr \st r \land
  %   \left(
  %     \begin{array}[c]{@{}l@{}}
  %       \left(\left(\ora{\reachable.Head.a \lor
  %             (\deref a\cdota mrk)}\right) \land \ora{\& n1_p = a}\right) \ch
  %       \qquad \\
  %       \hfill \Always ((\& n1_p = 
  %       a) \land \Sometime \neg (\deref a \cdota mrk) \land \Sometime (\deref a \cdota
  %       val = x))
  %     \end{array}\right)
  % }

  \trans{}{implies $\Always(a\mapsto mrk)$ in RHS, which contradicts $\Sometime
    \neg (a \mapsto mrk)$}

  \step{\exists a : Addr \st r \land \left(
      \begin{array}[c]{@{}l@{}}
        \left(% \left(
        \ora{\reachable.Head.a \land
          \neg (a\mapsto mrk) }% \right)
      % \land \ora{\& n1_p = a}
    \right)
    \ch 
    (\Always (n1_p = 
      a) \land \Sometime \neg (a \mapsto mrk) \land \Sometime ((a \mapsto
      val) = x))
    \end{array}\right)
  }
  % \step{\lor}
  % \step{\exists a : Addr \st r \land \left(\ora{\deref a\cdota mrk} \land \ora{\& n1_p = a}\right) \ch {} } 
  % \step{\hfill \Always ((\& n1_p = 
  %   a) \land \Sometime \neg (\deref a \cdota mrk) \land \Sometime (\deref a \cdota
  %   val = x))}

  \trans{\entails}{case analysis and assumption \refeq{eq:10},
    disjunct $\ora{(a\mapsto val) \neq x}$ in LHS of chop }

  \trans{}{implies $\Always((a\mapsto val) \neq x)$ in RHS, 
    contradicting $\Sometime ((a\mapsto val) = x)$}

  \step{\exists a : Addr \st r \land \left(
    \begin{array}[b]{@{}l@{}}
      % \left(
      \ora{inSet(Head,a,x)}%  \ora{\reachable.Head.a \land
          % \neg (a\mapsto mrk) \land ((a \mapsto
          % val) = x)}
      % \right)
      \ch {} 
       (\Always (n1_p = 
      a) \land \Sometime \neg (a \mapsto mrk) \land \Sometime ((a
      \mapsto 
      val) = x))
    \end{array}\right)
  }

  \trans{\entails}{definition of $absSet$}

  \step{\Sometime (x \in absSet)}

\end{derivation*}
Having shown that the behaviour of the implementation implies the
behaviour of the abstraction, it is now straightforward to show that
the refinement for case \refeq{eq:4} holds.

\smallskip

\noindent{\bf Proof of \refeq{eq:8}.} As with \refeq{eq:4}, we use
\refthm{thm:rely-enf-gc} to transfer the rely condition $r$ to the
right hand side as an enforced property. By logic, the right hand side
of the \refeq{eq:8} is equivalent to command $\enf r \land (\Always (x
\in absSet) \lor \Sometime(x \notin absSet)) \st CL \ch
clf_3:[\neg % (n1_p\mapsto mrk)
IN ]$.  The $\Sometime (x \notin absSet)$ case is trivially true. For
case $\Always (x \in absSet)$, we require that $r$ satisfies:
\begin{eqnarray}
& \Box 
(\Always (x \in absSet)  \imp 
  \exists a : Addr \st  
   \Always inSet(Head,a,x)
   )
\label{eq:11}
\\
& \Box(
\all k:\nat
\st \varphi^{k}.Head \neq Tail \imp
(\varphi^k.Head \mapsto val) < (\varphi^{k+1}.Head \mapsto val)
)
\label{eq:13}
\\
& 
\label{eq:24}
\Always (\reachable.n1_p.Tail)
\end{eqnarray}
By \refeq{eq:11}, in any interval, if the value $x$ is in the set
throughout the interval, there is an address that can be reached from
$Head$, the marked bit corresponding to the node at this address is
unmarked and the value field contains $x$. By \refeq{eq:13} the
reachable nodes of the list (including marked nodes) must be sorted in
strictly ascending order and by \refeq{eq:24} the $Tail$ node must be
reachable from $n1_p$. Conditions \refeq{eq:11}, \refeq{eq:13} and
\refeq{eq:24} together imply that there cannot be a terminating
execution of $\mathsf{CLoop}(p,x)$ such that $clf_3:[\neg IN ]$ holds,
i.e., the behaviour is equivalent to $false$.

\smallskip 
\noindent{\bf Proof of $S2$.}
The final rely condition $r$ must imply each of % \refeq{eq:14},
\refeq{eq:5}, \refeq{eq:10}, % \refeq{eq:1},
\refeq{eq:11}, \refeq{eq:13} and \refeq{eq:24}. We choose to take the
weakest possible instantiation and let $r$ be the conjunction
$% \refeq{eq:14} \land
\refeq{eq:5} \land \refeq{eq:10} \land % \refeq{eq:1} \land
\refeq{eq:11} \land \refeq{eq:13} \land \refeq{eq:24}$. These
properties are straightforward to verify by expanding the definitions
of the behaviours. The details of this proof are elided.

\section{Conclusions}

We have developed a framework, based on \cite{DDH12}, for reasoning
about the behaviour of a command over an interval that enables
reasoning about pointer-based programs where processes may refer to
states that are apparent to a process \cite{HBDJ13}. Parallel
composition is defined using conjunction and conflicting access to
shared state is disallowed using fractional permissions, which models
truly concurrent behaviour.  We formalise behaviour refinement in our
framework, which can be used to show that a fine-grained
implementation is a refinement of a coarse-grained abstraction. One is
only required to identify linearising statements of the abstraction
(as opposed to the implementation) and the proof of linearisability
itself is simplified due to the coarse-granularity of commands. For
the coarse-grained contains operation in \ref{fig:labs}, the guard
$\aang{x \in absSet}$ is the linearising statement for an execution
that returns $true$ and $\aang{x \notin absSet}$ the linearising statement of
an execution that returns $false$.

Our proof method is compositional (in the sense of rely/guarantee) and
in addition, we develop the rely conditions necessary to prove
correctness incrementally. As an example, we have shown refinement
between the \texttt{contains} operation of Heller et al's lazy set and
an abstraction of the contains operation that executes with
coarse-grained
atomicity. % Verification of the is in general considered difficult to
% verify
% \cite{VHHS06,CGLM06,DSW11,Vaf10}.
% Our semantic framework is based on
% , % which introduces fractional permissions as technique for
% % reasoning about interference. 
% % and methods for evaluating expressions non-deterministically (i.e.,
% % actual and apparent evaluation) are based on . Our notion
% % of behaviour refinement over intervals has been applied to both
% % concurrent \cite{DDH12} and real-time programs \cite{DH12MPC}, and has
% % has been used to prove linearisability of a simple stack algorithm
% % \cite{DD12}.
% % Note that an additional step is necessary complete the linearisability
% % proof for the lazy set \cite{DD12}. However, because we have shown
% % that the program in \reffig{fig:labs} is a coarse-grained abstraction
% % of ${\sf Contains}(p,x)$ in \reffig{fig:formal-lazyset}, the
% % linaerisability proof may be with respect to the simpler program in
% % \reffig{fig:labs}.

Behaviour refinement is defined in terms of implication, which makes
this work highly suited to mechanisation. However, we consider full
mechanisation to be future work. % Furthermore, the
% refinement theory can be used to develop fine-grained implementations
% from coarse-grained abstractions.

\noindent\textbf{Acknowledgements.}
  This research is supported by EPSRC Grant EP/J003727/1. We thank
  Gerhard Schellhorn and Bogdan Tofan for useful discussions, and
  anonymous reviewers for their insightful comments.

\bibliographystyle{plain}
\bibliography{thesis}

\newpage
\appendix

\section{Proofs of Add/Remove}
\newcommand{\spec}[1]{\left\lfloor #1 \right\rfloor}
\newcommand{\NE}[1]{\underline{ #1}}

In this appendix, we complete the proofs of abstraction for the
\texttt{add} and \texttt{remove} operations. Compared to the proofs of
the \texttt{contains} operation, these proofs are simpler due to the
locking that occurs during the main portion of each
operation. However, because we assume a truly concurrent semantics,
the coarse-grained abstraction is more difficult to specify. In
particular, it is possible for a number of concurrent add/remove
operations to take effect as part of a single state transition. 

% \begin{brijesh}
%   Mention the true concurrency aspect. 
% \end{brijesh}

\subsection{Formal model of \texttt{locate}, \texttt{add} and
  \texttt{remove}} 

In this section, we formalise the ${\sf Add}$ and ${\sf Remove}$
operations in our framework, which requires that we also formalise
${\sf Locate}$.
\begin{eqnarray*}
  located(pred,curr) & \sdef & \neg
  (pred\mapsto mrk) \land \neg (curr\mapsto mrk) \land
  ((pred\mapsto nxt) = curr) \\
  \mathsf{Search}(p, x, pred, curr) & \sdef & 
  \begin{array}[t]{@{}l@{}}
    pred \asgn Head \ch curr \asgn (pred \mapsto nxt) \ch \\
    ({[}(curr\mapsto val) < x{]}
    \ch pred \asgn curr \ch curr \asgn (pred\mapsto nxt))^\omega \ch \\
    {[}(curr\mapsto val)
    \geq x{]} \ch \mathsf{Lock}(p, pred) \ch \mathsf{Lock}(p, curr) \ch 
  \end{array}
  \\
  \mathsf{TryFind}(p, x, pred, curr) & \sdef &
  \begin{array}[t]{@{}l@{}}
    \mathsf{Search}(p, x, pred, curr) \ch 
    {[} \neg located(pred,curr) {]} \ch {} \\
    \mathsf{Unlock}(p, pred)
    \ch \mathsf{Unlock}(p, curr)
  \end{array}
  \\
  \mathsf{Find}(p, x, pred, curr) & \sdef &
  \begin{array}[t]{@{}l@{}}
    \mathsf{Search}(p, x, pred, curr) \ch 
    {[}located(pred,curr) {]} 
  \end{array}
  \\
  \mathsf{Locate}(p, x, pred, curr) & \sdef &
  \mathsf{TryFind}(p, x,pred,curr)^\omega \ch \mathsf{Find}(p, x,pred,curr)
\end{eqnarray*}
We define a predicate $located(pred, curr)$, which formalises the
guard at \texttt{L8}. Operation 
$$
\mathsf{Search}(p, x, pred, curr)
$$ 
formalises lines \texttt{L1}-\texttt{L7} and $\mathsf{TryFind}(p, x,
pred, curr)$ formalises an execution of \texttt{L8} in which guard
$located(pred, curr)$ evaluates to $false$. The two unlock statements
within $\mathsf{TryFind}(p, x, pred, curr)$ correspond to \texttt{L10}
and \texttt{L11}. The $\mathsf{TryFind}(p, x, pred, curr)$ operation
models an execution of the main loop body within \texttt{locate} that
that loops again. Operation $\mathsf{Find}(p, x, pred, curr)$ models a
successful execution of the loop body (where $located(pred, curr)$
evaluates to $true$.
$$
\begin{array}[t]{l@{\qquad}l@{\qquad}l}
  \begin{array}[t]{@{}l@{}}
    \mathsf{AddOK}(p,x)  \sdef\\
    alt_2:  [(n3_p\mapsto val) \neq x] \ch {} \\
    al_3: NewNode(x, n2_p) \ch  {} \\
    al_4: (n2_p \cdota nxt) \hasgn n3_p \ch {} \\ 
    al_5: (n1_p \cdota nxt) \hasgn n2_p \ch  {} \\
    al_6: res_p \asgn true
  \end{array}
  & 
  \begin{array}[t]{@{}l@{}}
    \mathsf{AddFail}(p,x) \sdef \\
    alf_2: [(n3_p\mapsto val) = x] \ch {} \\
    al_7: res_p \asgn
    false
  \end{array}
  & 
  \begin{array}[t]{@{}l@{}}
    \mathsf{Add}(p,x) \sdef \\
    al_1: \mathsf{Locate}(p, x, n1_p, n3_p) \ch {} \\
    (\mathsf{AddOK}(p,x)
    \sqcap \mathsf{AddFail}(p,x)) \ch {} \\
    al_8:  \mathsf{Unlock}(p, n1_p) \ch {}\\
    al_9:  \mathsf{Unlock}(p, n3_p)
  \end{array}
\end{array}
$$
As the names imply, $\mathsf{AddOK}$ and $\mathsf{AddFail}$ model the
successful and failed executions of the add operation, and
$\mathsf{Add}$ operation behaves as locate, then non-deterministically
chooses between an successful or failed operation, then unlocks the
locks on $n1_p$ and $n3_p$ held after the termination of
$\mathsf{Locate}$. Operation $Remove$ is similar, and is formalised
below. 
$$
\begin{array}[t]{l@{\qquad}l@{\qquad}l}
  \begin{array}[t]{@{}l@{}}
    \mathsf{RemOK}(p,x) \sdef \\
    rlt_2: [(n2_p\mapsto val) = x] \ch {} \\
    rl_3: (n2_p \cdota mrk) \hasgn true
    \ch {} \\
    rl_4: n3_p \asgn (n2_p\mapsto nxt) \ch {} \\
    rl_5: (n1_p \cdota nxt) \hasgn n3_p \ch {} \\
    rl_6: res_p \asgn true
  \end{array}
  & 
  \begin{array}[t]{@{}l@{}}
    \mathsf{RemFail}(p,x)  \sdef \\
    rlf_2: [(n2_p\mapsto val) \neq x]
    \ch {} \\
    rl_7: res_p \asgn false
  \end{array}
  & 
  \begin{array}[t]{@{}l@{}}
    \mathsf{Remove}(p,x) \sdef \\
    rl_1: \mathsf{Locate}(p,x, n1_p, n2_p) \ch {} \\
    (\mathsf{RemOK}(p,x)\sqcap \mathsf{RemFail}(p,x)) \ch {} \\
    rl_3: \mathsf{Unlock}(p, n1_p) \ch {} \\
    rl_4: \mathsf{Unlock}(p, n2_p)
  \end{array}
\end{array}
$$

\subsection{The \texttt{add} operation}

In this section, we verify the coarse-grained abstraction of the
\texttt{add} operation.  Unlike the \texttt{contains} operation, this
abstraction cannot be defined using the standard language constructs,
because the standard constructs are not precise enough to describe the
abstract behaviour.  Hence, we introduce a specification command,
which turns an interval predicate to into a command, whose behaviour
is given by the interval predicate. Thus, for an interval predicate
$g$, process $p$ and set of variables $Z$, the behaviour of a
specification command is given by:
\begin{eqnarray*}
  beh_{p,Z}.\spec{g} & \sdef & g
  % beh_{p,Z}.\Chaos & \sdef & true
\end{eqnarray*}

We also introduce two further interval predicates, namely $\NE{g}$,
which states that interval predicate $g$ holds and the interval under
consideration is non-empty, and $\Diamond g$ which states that $g$
holds in some subinterval of the given interval, i.e., for an interval
$\Delta$ and stream $s$, we define: 
\begin{eqnarray*}
  \NE{g}
  & \sdef &  \neg \Empty \land g \\
  (\Diamond g).\Delta.s & \sdef & \exists \Delta' : Interval \st
  \Delta' \subseteq \Delta \land g.\Delta'.s
\end{eqnarray*}
We define a state predicate $WriteFields(p, a, F)$ which holds if
process $p$ writes to any of the fields in $F$ of the data structure
at address $a$.
\begin{eqnarray*}
  WriteFields(p, a, F) & \sdef & \exists b : \{a 
  \cdota f \mid f \in F\} \st 
  \mcW.b.p
\end{eqnarray*}
We further define interval predicate $ModSet.p$ that is used to
determine whether $p$ ever writes to the addresses corresponding to
the $val$, $mrk$ and $nxt$ fields of the nodes reachable from $Head$,
$IntFree(p,n)$, which holds if no other process different from $p$
writes to fields of the node $n$, and $Insert(p, x)$ that restricts the
values that are modified by $p$ with respect to node $n$.
\begin{eqnarray*}
ModSet.p & \sdef & 
\Sometime \exists a :  setAddr \st WriteFields(p,a,\{ val,  mrk,
nxt\})
\\
IntFree(p, n) & \sdef & \Always \neg \mcI.\{n \cdota val, n \cdota mrk, 
n \cdota nxt, n \cdota lck\}.p 
\end{eqnarray*}
Thus, $ModSet.p$ holds iff there is a point in the interval such that
$p$ writes to the $val$, $mrk$ or $nxt$ fields of the node at address
$a$ and $IntFree(p, n)$ holds iff there is no interference by the
environment of $p$ to any of the fields of node $n$.

The insertion of a node into the set is modelled as follows, where
$preIns$ denotes the precondition of an insertion, $doIns$ models the
insertion, and $Insert$ models the full operation, including the
possible interference from other processors. 
\begin{eqnarray*}
preIns(a, b, x) & \sdef & \reachable.Head.a \land located(a,b) \land (a \mapsto val <
  x) \land (b \mapsto val > x)
\\
doIns(a, n, b, x) & \sdef & (a \mapsto nxt = n) \land (n \longmapsto (x,
b, false, null))
\\
Insert(p, x) & \sdef &
\exists a, n, b : Addr \st
\begin{array}[t]{@{}l@{}}
\prev\ \ora{preIns(a,b,x)} \land  \NE{\Always doIns(a,n,b,x)}\land\\
IntFree.a \land IntFree.b \land 
 \all ua : Addr \bs \{a\cdota nxt\} \st \Always \neg \mcW.ua.p  
\end{array}
% \Always \left(
%   \begin{array}[c]{@{}l@{}}
%     (\all v : Var \st \neg \mcW.v.p) \land \\
%     \left(
%       \begin{array}[c]{@{}l@{}}
%         \all a : Addr \st
%         \begin{array}[t]{@{}l@{}}
%           (\mcW.a.p
%           \imp a \cdota nxt = n \land \reachable.Head.a) \land \\
%           (a \cdota nxt = n
%           \land \reachable.Head.a \imp (a \mapsto nxt = n))
%         \end{array}
%       \end{array}
%     \right)
%   \end{array}
% \right)
% \all a : Addr \st
% \Always 
% \left(
%     \begin{array}[c]{@{}l@{}}
%       \reachable.Head.a \imp \\
%       \quad \kif\ (a \mapsto nxt) \neq n \\
%       \quad \kthen\ 
%       \neg WriteFields(p, a, \{val,nxt,lck,mrk\})  
%     % \begin{array}[t]{@{}l@{}}
%     %     \neg \mcW.(a\cdota val).p \land \neg
%     %     \mcW.(a\cdota nxt).p  \land \\
%     %     \neg \mcW.(a\cdota lck).p \land \neg
%     %     \mcW.(a\cdota mrk).p
%     %   \end{array}
%       \\
%       \quad \kelse\ \neg WriteFields(p, a, \{val,lck,mrk\})  
%     %   \begin{array}[t]{@{}l@{}}
%     %   \neg \mcW.(a\cdota val).p \land \neg \mcW.(a \cdota lck).p \land\\
%     %   \neg \mcW.(a \cdota mrk).p 
%     % \end{array}
%     \end{array}
%   \right)
\end{eqnarray*}
State predicate $preIns(a, b, x)$ states that $a$ is reachable from
$Head$, the $located(a,b)$ predicate holds, node $a$ has value is less
than $x$ and node $b$ has value greater than $x$. Thus, $x$ is not in
the abstract set. State predicate $doIns(a, n, b, x)$ states that $a
\cdot nxt$ is updated with value $n$, and node $n$ has value $x$,
points to $b$ is not marked and is not locked. The $Insert(p, x)$
predicate states that there are addresses $a$, $n$ and $b$ such that
$preIns(a,b,x)$ holds as a precondition, behaves as $doIns(a,n,b,x)$
and furthermore, $a$ and $b$ are interference free and $p$ does not
write to any other set address.

The coarse-grained abstraction of the \texttt{add} operation is then
defined as follows.
\begin{eqnarray*}
  \mathsf{CGAOK}(p, x) & \sdef &
  \begin{array}[t]{@{}l@{}}
    % \spec{\neg ModSet.p}\ch 
    \spec{
    % \Diamond
    % \left(
    \begin{array}[c]{@{}l@{}}
         Insert(p, x)%  \land  \\
      % \Always \left(inSet(n2_p,x) \land
      % (\ref{eq:13}) \land (\ref{eq:24})
    \end{array}
    % \right)
  } \ch 
  res_p \asgn true 
\end{array}
  \\
  % \mathsf{CGAFail}(p, x) & \sdef & \Enf{\Always \neg
  %   ModSet.p
  %   % inSet(Head, n2_p, x)
  % }{\spec{\Diamond\NE{\Always (x \in S)}} \ch 
  %   res_p \asgn false} 
  \mathsf{CGAFail}(p, x) & \sdef & \AG{x \in absSet} \ch 
    res_p \asgn false
  \\
  \mathsf{CGAdd}(p, x) & \sdef &  \spec{\neg ModSet.p} \ch 
  (\mathsf{CGAOK}(p, x) \sqcap \mathsf{CGAFail}(p, x)) \ch \spec{\neg ModSet.p}
\end{eqnarray*}
A successful execution of the \texttt{add} operation behaves as
$Insert(p,x)$ then sets the return value $res_p$ to $true$. A failed
execution of the \texttt{add} operation never adds $n2_p$ to the set,
but detects that $x$ is in the set and sets $res_p$ to $false$. The
$\mathsf{Add}$ operation performs some idling at the start modelled as
$\neg ModSet.p$ because the concrete operation has the possibility of
not terminating, and at the end (to allow the concrete program time to
unlock the held locks).

Like the decomposition depicted in \reffig{fig:refsteps} for the
\texttt{contains} operation, we may decompose the proof so that we
consider the execution of \texttt{add} by a single process under a
rely condition $r$ that we assume splits, provided that the rest of
the program satisfies the rely condition that we derive. Given that
$\mathsf{CGS}'$ is the program derived from $\mathsf{CGS}$ by
replacing $\mathsf{Add}$ by $\mathsf{CGAdd}$, the refinement holds if
we prove both of the following:
\begin{eqnarray}
  \label{eq:22}
  \Rely{r}{\mathsf{CGS'}(p)} & \sref_{p}^{HT}  & \mathsf{S}(p) 
  \\
  \label{eq:23}
  beh_{Q,HT}.(\Par_{q : Q} \mathsf{S}(q))  & \entails &  r
\end{eqnarray}

\subsubsection{Proof of (\ref{eq:22}).}

We define the following state predicate, which formalises the
postcondition of $\mathsf{Locate}(p,pred,curr)$.
\begin{eqnarray*}
  postLocate(p,pred,curr) & \sdef & 
  \begin{array}[t]{@{}l@{}}
    located(pred,curr) \land \\
    ((pred \mapsto val) < x) \land ((curr \mapsto val) \geq x) \land \\
     ((curr \mapsto lck) = p)
    \land ((pred \mapsto
    lck) = p) \land \\
    \reachable.Head.pred \land \reachable.Head.curr
  \end{array}
\end{eqnarray*}
Thus, operation {\sf Locate} ensures that $pred$ and $curr$ satisfy
$located$, that the value of $pred$ is less than $x$, the value of
$curr$ is above or equal to $x$, that both $curr$ and $pred$ are
locked, and that both $pred$ and $curr$ are reachable from $Head$.  We
now have the following refinement, where $\mathsf{U}(p, n_1, n_2)
\sdef \mathsf{Unlock}(p, n_1) \ch \mathsf{Unlock}(p, n_2)$.
\begin{derivation}
  \step{\Enf{r}{\mathsf{Add}(p,x)}}

  \trans{\srefeq_p^M}{definition of $\mathsf{Add}(p,x)$}

  \step{\enf r \st \mathsf{Locate}(p,x,n1_p, n3_p) \ch ({\sf AddOK}(p,x) \sqcap {\sf
      AddFail}(p,x)) \ch {\sf U}(p,n1_p, n3_p)}

  \trans{\srefeq_p^M}{logic}

  \step{\enf r \st
    \begin{array}[t]{@{}l@{}}
      \left(\Enf{\Inf}{\mathsf{Locate}(p,x,n1_p, n3_p)}) \sqcap
        (\Enf{\Fin}{\mathsf{Locate}(p,x,n1_p, n3_p)})\right) \ch {} 
      \\
      ({\sf AddOK}(p,x) \sqcap {\sf
        AddFail}(p,x)) \ch {\sf U}(p,n1_p, n3_p)
    \end{array}
  }

  \trans{\sqsupseteq_p^M}{behaviour of $\mathsf{Locate}(p,x,n1_p, n3_p)$}
  \trans{}{distribute `$\sqcap$' over `;', $\Inf$ is a right
    annihilator}

  \step{\enf r \st \spec{\Inf \land \neg ModSet.p} \sqcap {}}
  \step{\enf r \st
    \begin{array}[t]{@{}l@{}}
      (\Enf{\Fin}{\mathsf{Locate}(p,x,n1_p, n3_p)}) \ch
      {} \\
      ({\sf AddOK}(p,x) \sqcap  {\sf AddFail}(p,x)) \ch {\sf U}(p,n1_p,n3_p)
    \end{array}
  }
  
  \trans{\srefeq_p^M}{distribute `$\sqcap$'}
  \step{\enf r \st \spec{\Inf \land \neg ModSet.p} \sqcap {}
    \hfill (A_1)}

  \step{\enf r \st (\Enf{\Fin}{\mathsf{Locate}(p,x,n1_p, n3_p)}) \ch {\sf AddOK}(p,x) \ch {\sf
      U}(p,n1_p,n3_p) \sqcap {} \hfill (A_2)}
   
  \step{\enf r \st (\Enf{\Fin}{\mathsf{Locate}(p,x,n1_p, n3_p)}) \ch {\sf AddFail}(p,x) \ch {\sf
      U}(p,n1_p,n3_p) \hfill (A_3)}
  
\end{derivation}

Splitting the behaviour of $\mathsf{Locate}(p,x,n1_p, n3_p)$ into finite and
infinite executions, and distributing the `;' through `$\sqcap$', it
is possible to show that $\mathsf{CGAdd}(p, x) \srefeq_p^L (CA_1) \sqcap (CA_2) \sqcap (CA_3)$
where $(CA_1)$,  $(CA_2)$, and  $(CA_3)$ are defined below. 
\begin{derivation*}
  \step{\spec{\Inf \land \neg ModSet.p} \hfill (CA_1)}

  \step{\spec{\Fin \land \neg ModSet.p}  \ch \mathsf{CGAOK}(p, x) \ch
    \spec{\neg ModSet.p} \hfill (CA_2)}
  
  \step{\spec{\Fin \land \neg ModSet.p}  \ch  \mathsf{CGAFail}(p, x) \ch
    \spec{\neg ModSet.p} \hfill (CA_3)}
\end{derivation*}

\paragraph{Proof of $(A_1)$.}
It is trivial to verify $(CA_1) \sref_p^{L,M} (A_1)$. 

\paragraph{Proof of $(A_2)$.} To prove this case, we strengthen
condition $r$ and require that it satisfies both of the following.
\begin{eqnarray}
  \label{eq:21}
  r & \entails & \Box (\Always(pc_p \in \{al_i \mid i \in
  [2,7]\}) \imp IntFree.n1_p \land IntFree.n3_p)
  \\
  \label{eq:20}
  r & \entails & \Always (pc_p \in \{al_4,al_5\}) \imp
  stable.\{n2_p\cdota val, n2_p\cdota mrk, n2_p\cdota nxt\}
\end{eqnarray}
Assuming that $r$ holds, we now have the following calculation.
\begin{derivation*}
  \step{beh_{p,M}.
    \left(\Enf{\Fin}{\mathsf{Locate}(p,x,n1_p, n3_p)} \ch {\sf AddOK}(p,x) \ch {\sf
        U}(p,n1_p,n3_p)\right)}

  % \trans{\equiv}{expand behaviour}

  % \step{beh_{p,M}.(\Enf{\Fin}{\mathsf{Locate}(p,x)} ) \ch beh_{p,M}.({\sf
  %     AddOK}(p,x) \ch beh_{p,M}.{\sf U}(p,n1_p,n3_p))}

  \trans{\entails}{expand behaviour and use \refeq{eq:21} }

  \step{beh_{p,M}.(\Enf{\Fin}{\mathsf{Locate}(p,x,n1_p, n3_p)}) \ch{}} 
  
  \step{(IntFree.n1_p \land IntFree.n3_p \land beh_{p,M}.{\sf
      AddOK}(p,x)) \ch {} }

  \step{beh_{p,M}.{\sf U}(p,n1_p,n3_p)}

  \trans{\entails}{first chop: behaviour of $\mathsf{Locate}$}
  \trans{}{second chop: expand $\mathsf{AddOk}(p,x)$, use
    $postLocate$, behaviour of $alt_2$ and $IntFree$ conditions}

  \step{
    (\neg ModSet.p \land \ora{postLocate(p,n1_p,n3_p)})
    \ch{}}
  \step{
    \left(\begin{array}[c]{@{}l@{}}
        \exists 
        a: Addr, k:Val \st 
          \begin{array}[t]{@{}l@{}}
            \left(
              \begin{array}[c]{@{}l@{}}
                \Always (n1_p \mapsto
                val < x) \land \Always(n3_p \mapsto val > x) \land\\
                \left(
                  \begin{array}[c]{@{}l@{}}
                    beh_{p,M}.(alt_2 \ch al_3 \ch al_4)
                    \ch \\
                    \Always(pc_p = al_5) \land \Eval_{p,M}.(n1_p \cdota nxt = a
                      \land n2_p = k)
                  \end{array}
                \right)
              \end{array}
            \right) \ch \!\!
            \\
            (\Always(pc_p =
            al_5) \land \Update_{p,M}.(a,k))
          \end{array}
        \end{array}\right)
    \ch {}}

  \step{ beh_{p,M}.al_6 \ch beh_{p,M}.{\sf U}(p,n1_p,n3_p)}

  \trans{\entails}{first chop: logic, second chop: expand behaviour
    use (\ref{eq:20})} 

  \step{ \neg ModSet.p
    \ch{}}
  \step{
    \left(\begin{array}[c]{@{}l@{}}
        \exists 
        a: Addr, k:Val \st 
          \begin{array}[t]{@{}l@{}}
            \left(
              \begin{array}[c]{@{}l@{}}
                    \neg ModSet.p \land \ora{preIns(n1_p, n3_p, x)}
                    \land (\ora{n2_p \mapsto (x, n3_p, false,null)}) %beh_{p,M}.(alt_2 \ch al_3 \ch al_4)
                    \land \\
                    \ora{(n1_p \cdota nxt = a
                      \land n2_p = k)}
              \end{array}
            \right)
            \ch \!\!
            \\
            (\Always(pc_p = al_5) \land \Update_{p,M}.(a,k))
          \end{array}
        \end{array}\right)
    \ch {}}

  \step{ beh_{p,M}.al_6 \ch beh_{p,M}.{\sf U}(p,n1_p,n3_p)}

  \trans{\equiv}{logic, $\neg ModSet.p$ both splits and joins}

  \step{
    \neg ModSet.p
    \ch{}}
  \step{
    \left(\begin{array}[c]{@{}l@{}}
        \exists 
        a: Addr, k:Val \st 
          \begin{array}[t]{@{}l@{}}
            \prev\left(
              \begin{array}[c]{@{}l@{}}
                    \ora{preIns(n1_p, n3_p, x)}
                    \land (\ora{n2_p \mapsto (x, n3_p, false,null)}) %beh_{p,M}.(alt_2 \ch al_3 \ch al_4)
                    \land \\
                    \ora{(n1_p \cdota nxt = a
                      \land n2_p = k)}
              \end{array}
            \right)
            \land
            \\
            (\Always(pc_p = al_5) \land \Update_{p,M}.(a,k))
          \end{array}
        \end{array}\right)
    \ch {}}

  \step{ beh_{p,M}.al_6 \ch beh_{p,M}.{\sf U}(p,n1_p,n3_p)}

  \trans{\entails}{$p$ holds locks on $n1_p$ and $n3_p$, use
    (\ref{eq:20}), definition of $\Update$}

  \step{
    \neg ModSet.p
    \ch{}}
  \step{
    \left(\begin{array}[c]{@{}l@{}}
            \prev\ \ora{preIns(n1_p, n3_p, x)}
            \land 
            \NE{\Always\ doIns(n1_p, n2_p, n3_p, x)} \land \\
            IntFree.n1_p \land IntFree.n3_p \land 
            \all ua : Addr \bs \{a\cdota nxt\} \st \Always \neg \mcW.ua.p  
        \end{array}\right)
    \ch {}}

  \step{ beh_{p,M}.al_6 \ch beh_{p,M}.{\sf U}(p,n1_p,n3_p)}

  \trans{\entails}{change context}

  \step{beh_{p,L}.(CA_2)}%\mathsf{CGAOK}(p,x)}

\end{derivation*}

\paragraph{Proof of $(A_3)$.} Once again assuming $r$ holds, we obtain: 
\begin{derivation*}
  \step{beh_{p,M}.\left(\Enf{\Fin}{\mathsf{Locate}(p,x,n1_p, n3_p)} \ch {\sf
        AddFail}(p,x) \ch {\sf U}(p,n1_p,n3_p)\right)}

  \trans{\entails}{definition of $beh$ and behaviour of
    $\mathsf{Locate}$}

  \step{(\neg ModSet.p \land \ora{postLocate(p,n1_p, n3_p)}) \ch beh_{p,M}.{\sf AddFail}(p,x) \ch beh_{p,M}.{\sf U}(p,n1_p,n3_p)}

  \trans{\entails}{definition of  $postLocate(p,n1_p, n3_p)$}

  \step{\left(
    \begin{array}[c]{@{}l@{}}
      beh_{p,L}.\Idle \land\\ 
      \ora{\reachable.Head.n3_p \land \neg (n3_p \mapsto
        mrk)}
    \end{array}
  \right) \ch  beh_{p,M}.{\sf AddFail}(p,x) 
    \ch beh_{p,M}.{\sf U}(p,n1_p,n3_p)}

  \trans{\entails}{use \refeq{eq:21} and guard $alf_3$, change context}

  \step{beh_{p,L}.\Idle   \ch  (beh_{p,L}.\Idle \land \NE{\Always (\reachable.Head.n3_p
      \land \neg (n3_p \mapsto mrk) \land (n3_p \mapsto val = x))})
    \ch {} }
  \step{beh_{p,M}.(al_7 \ch {\sf U}(p,n1_p,n3_p))}

  \trans{\entails}{definition of $absSet$}

  \step{beh_{p,L}.\Idle
    \ch (beh_{p,L}.\Idle \land \NE{\Always (x \in absSet)})
    \ch  beh_{p,M}.(al_7 \ch {\sf U}(p,n1_p,n3_p))}

  \trans{\entails}{change context}

  \step{beh_{p,M}.CA_3}%{\sf CGAFail}(p,x)}

\end{derivation*}

\subsubsection{Proof of (\ref{eq:23}).}

We strengthen the rely condition with additional conjunct
$\refeq{eq:21}\land\refeq{eq:20}$. This proof is straightforward due
to the locks held by process $p$.

\subsection{The \texttt{remove} operation}

A coarse-grained abstraction of the \texttt{remove} operation is given
below. 
\begin{eqnarray*}
  preDel(p, a, n, b, x) & \sdef &
  \begin{array}[t]{@{}l@{}}
    \reachable.Head.a \land located(a,n) \land (a \mapsto val <
    x) \land \\
    (n \longmapsto (x, b, false, p))
  \end{array}
  \\
  doDel(a, n, b, x) & \sdef & \Always ((a \mapsto
      nxt = b) \lor (n \mapsto mrk))
  \\
  Delete(p, x) & \sdef &
  \exists a, n, b : Addr \st 
  \begin{array}[t]{@{}l@{}}
    \prev \ora{preDel(p, a, n, b, x)} \land 
    \NE{\Always doDel(a, n, b, x)}\land \\
    IntFree.a \land IntFree.n \land \\
    \all ua : Addr \bs \{a\cdota nxt, n \cdota mrk\} \st \Always \neg \mcW.ua.p  
  \end{array}
  \\
  \mathsf{CGROK}(p, x) & \sdef &
  \begin{array}[t]{@{}l@{}}
    \spec{
      % \Diamond
      % \left(
      \begin{array}[c]{@{}l@{}}
        % \NE{\Always (x \notin 
        % S)} \ch
        Delete(p,x)
        % \begin{array}[c]{@{}l@{}}
        %   \reachable.Head.n2_p \land \\
        %   \neg (n2_p \mapsto
        %   mrk) \land (n2_p \mapsto val = x)
        % \end{array}
      \end{array}
      % \right)
    } \ch 
    res_p \asgn true 
  \end{array}
  \\
  \mathsf{CGRFail}(p, x) & \sdef & 
  % \left(\Enf{\Always \neg
  %   ModSet.p}{\spec{\Diamond\NE{\Always (x \notin S)}}}\right) \ch 
  \AG{x \notin absSet} \ch
  res_p \asgn false 
  \\
  \mathsf{CGR}(p, x) & \sdef & \spec{\neg ModSet.p} \ch (\mathsf{CGROK}(p, x) \sqcap \mathsf{CGRFail}(p, x))\ch \spec{\neg ModSet.p}
\end{eqnarray*}
The proof of refinement between \texttt{remove} and the abstraction
above proceeds in a similar manner to the \texttt{add} operation.
In particular, $\mathsf{CGR}(p, x) \sref_p^L (CR_1) \sqcap (CR_2) \sqcap
(CR_3)$ holds, where:
\begin{derivation*}
  \step{\spec{\Inf \land \neg ModSet.p}\hfill (CR_1)}
  
  \step{\spec{\Fin \land \neg ModSet.p} \ch \mathsf{CGROK}(p, x) \ch \spec{\neg ModSet.p}\hfill (CR_2)}
  
  \step{\spec{\Fin \land \neg ModSet.p} \ch \mathsf{CGRFail}(p, x) \ch
    \spec{\neg ModSet.p}\hfill (CR_3)}
\end{derivation*}
Furthermore, $(R_1) \sqcap (R_2) \sqcap (R_3)\sref_p^{L,M}
\mathsf{Remove}(p,x)$ holds, where:
\begin{derivation*}
  \step{(\Enf{\Inf}{\mathsf{Locate}(p,x, n1_p, n2_p)}) \hfill (R_1) }

  \step{(\Enf{\Fin}{\mathsf{Locate}(p,x, n1_p, n2_p)}) \ch
    \mathsf{RemOK}(p,x) \ch \mathsf{U}(p, n1_p, n2_p)\hfill (R_2)}

  \step{(\Enf{\Fin}{\mathsf{Locate}(p,x, n1_p, n2_p)}) \ch
    \mathsf{RemFail}(p,x) \ch \mathsf{U}(p, n1_p, n2_p) \hfill (R_3)}
\end{derivation*}
Thus, to complete the proof, we must show: $(CR_i) \sref_p^{L,M}
(R_i)$ for $i \in \{1,2,3\}$, and the proof of $i = 3$ is similar to
the failed case of the {\tt add}. The proof of $i = 1$ is trivial. For the
proof of case $i = 2$ we assume the following.
\begin{eqnarray}
  \label{eq:27}
  r & \entails & \Box (\Always(pc_p \in \{rl_i \mid i \in
  [2,7]\}) \imp IntFree.n1_p \land IntFree.n2_p)
  % \\
  % \label{eq:20}
  % r & \entails & \Always (pc_p \in \{al_4,al_5\}) \imp
  % stable.\{n2_p\cdota val, n2_p\cdota mrk, n2_p\cdota nxt\}
\end{eqnarray}
Hence, assuming $r$, the proof proceeds as follows.
\begin{derivation*}
  \step{beh_{p,M}.(R_2)}
  
  \trans{\equiv}{expanding definitions}

  \step{beh_{p,M}.(\Enf{\Fin}{\mathsf{Locate}(p,x, n1_p, n2_p)}) \ch  {}}

  \step{\left(
      \begin{array}[c]{@{}l@{}}
        \exists a: Addr, k : Val \st
        \begin{array}[t]{@{}l@{}}
          beh_{p,M}.rlt_2 \ch (\Always (pc_p = rl_3) \land \Eval_{p,M}.(k \land
          (a = (n2_p \cdota mrk)))) \ch {} \\
          (\Always (pc_p = rl_3) \land \Update_{p,M}.(a, k)) \ch beh_{p,M}.(rl_4 \ch rl_5
          \ch rl_6)
        \end{array}
      \end{array}
    \right)  \ch {}}

  \step{beh_{p,M}.{\sf U}(p, n1_p, n2_p)}

  \trans{\entails}{behaviour of {\sf Locate}, then assuming (\ref{eq:27})}

  \step{(\neg ModSet.p \land \ora{postLocate(p, n1_p, n2_p)}) \ch  {}}

  \step{\left(
      \begin{array}[c]{@{}l@{}}
        \exists a: Addr, k : Val \st
        \begin{array}[t]{@{}l@{}}
          beh_{p,M}.rlt_2 \ch (\Always (pc_p = rl_3) \land \Eval_{p,M}.(k \land
          a = (n2_p \cdota mrk))) \ch {} \\
          (\Always (pc_p = rl_3) \land \Update_{p,M}.(a,  k)) \ch beh_{p,M}.(rl_4 \ch rl_5
          \ch rl_6)
        \end{array}
      \end{array}
    \right)  \ch {}}

  \step{beh_{p,M}.{\sf U}(p, n1_p, n2_p)}

  \trans{\entails}{logic, expand behaviours, use (\ref{eq:27})}

  \step{\neg ModSet.p \ch  {}}

  \step{\left(
      \begin{array}[c]{@{}l@{}}
        \exists b: Addr \st
        \begin{array}[t]{@{}l@{}}
          \neg ModSet.p \land  \ora{postLocate(p, n1_p, n2_p)} \land
          \ora{(n2_p \longmapsto (x, b, false, p))} \ch {} \\
          (\Always (pc_p = rl_3) \land \Update_{p,M}.(n2_p \cdota mrk, true)) \ch beh_{p,M}.(rl_4 \ch rl_5
          \ch rl_6)
        \end{array}
      \end{array}
    \right)  \ch {}}

  \step{beh_{p,M}.{\sf U}(p, n1_p, n2_p)}

  \trans{\equiv}{$\neg ModSet.p$ splits and joins, logic}

  \step{\neg ModSet.p \ch  {}}

  \step{\left(
      \begin{array}[c]{@{}l@{}}
        \exists b: Addr \st
        \begin{array}[t]{@{}l@{}}
          \prev(\ora{postLocate(p, n1_p, n2_p)} \land
          \ora{(n2_p \longmapsto (x, b, false, p))}) \land \\
          (\Always (pc_p = rl_3) \land \Update_{p,M}.(n2_p \cdota mrk, true)) \ch beh_{p,M}.(rl_4 \ch rl_5
          \ch rl_6)
        \end{array}
      \end{array}
    \right)  \ch {}}

  \step{beh_{p,M}.{\sf U}(p, n1_p, n2_p)}

  \trans{\entails}{behaviour definitions, (\ref{eq:27})}

  \step{beh_{p,M}.(CR_2)}
\end{derivation*}

\noindent
Finally, we are left with a proof requirement that the rest of the
program implies the rely condition \refeq{eq:27}. This proof is
straightforward due to the locks held by process $p$. 

\end{document}